\documentclass[journal,10pt]{IEEEtran}

\usepackage{graphicx,times,amsmath,booktabs,algorithm,algorithmic,amsmath}
\usepackage{multicol}
\usepackage{amsmath,amssymb}
\usepackage{subfigure}
\usepackage{stfloats}
\usepackage{amsfonts,bm}
\usepackage{mathrsfs}
\usepackage{bbm}

\usepackage{epstopdf}
\usepackage{cite}

\usepackage{color}

\usepackage{graphicx}
\usepackage{multirow}
\usepackage[table,xcdraw]{xcolor}
\usepackage{hyperref}

\begin{document}
\title{CaFTRA: Frequency-Domain Correlation-Aware Feedback-Free MIMO Transmission and Resource Allocation for 6G and Beyond}


\author{Bo~Qian,~\IEEEmembership{Member,~IEEE},
        Hanlin~Wu,~\IEEEmembership{Student Member,~IEEE},
        Jiacheng~Chen,~\IEEEmembership{Member,~IEEE},
        Yunting~Xu,~\IEEEmembership{Member,~IEEE},
        Xiaoyu~Wang,
        Haibo~Zhou, \IEEEmembership{Fellow,~IEEE},
        Yusheng~Ji,~\IEEEmembership{Fellow,~IEEE}

\thanks{This article has been accepted for publication in
IEEE Transactions on Mobile Computing.
\copyright\ 2026 IEEE. Personal use of this material is permitted.
Permission from IEEE must be obtained for all other uses, in any
current or future media, including reprinting/republishing this
material for advertising or promotional purposes, creating new
collective works, for resale or redistribution to servers or lists,
or reuse of any copyrighted component of this work in other works.
DOI: \protect\href{https://doi.org/10.1109/TMC.2026.3696984}{10.1109/TMC.2026.3696984}}
\thanks{This work was supported by in part by the Guangdong S\&T Programme under Grant 2024B0101010003, in part by the JSPS Grant-in-Aid for Early-Career Scientists under Grant JP25K21195, in part by the JSPS KAKENHI under Grant JP24K02937, in part by the JST ASPIRE under Grant JPMJAP2325, in part by the
National Key R\&D Program of China under Grant 2024YFE0200801 and Grant 2024YFE0200804, and in part by Mobile Information Networks-National Science and Technology Major Project under Grant 2025ZD1304900. \emph{(Corresponding author: Haibo Zhou.)}}
\thanks{Bo Qian and Hanlin Wu are with the Graduate School of Information Science and Technology, The University of Tokyo, Tokyo 113-8657, Japan (e-mail: boqian@ieee.org; hanlinwu@g.ecc.u-tokyo.ac.jp).}
\thanks{Jiacheng Chen is with the Department of Strategic and Advanced Interdisciplinary Research, Peng Cheng Laboratory, Shenzhen 518066, China (e-mail: chenjch02@pcl.ac.cn).}
\thanks{Yunting Xu is with the College of Computing and Data Science, Nanyang Technological University, Singapore (email: yunting.xu@ntu.edu.sg).}
\thanks{Xiaoyu Wang is with the Department of Informatics, Graduate University for Advanced Studies, SOKENDAI and the Information Systems Architecture Science Research Division, National Institute of Informatics, Tokyo 101-8430, Japan (e-mail: wangxiaoyu@nii.ac.jp).}
\thanks{Haibo Zhou is with the School of Electronic Science and Engineering, Nanjing University, Nanjing 210023, China (e-mail: haibozhou@nju.edu.cn).}
\thanks{Yusheng Ji is with National Institute of Informatics, Tokyo 101-8430, Japan (e-mail: kei@nii.ac.jp).}
}

\maketitle
\begin{abstract}
The fundamental designs of wireless systems toward AI-Native 6G and beyond are driven by the need for ever-increasing demand of mobile data traffic, extreme spectral efficiency and adaptability across diverse service scenarios.
To overcome the limitations posed by feedback-based multiple-input and multiple-output (MIMO) transmission, we propose a novel frequency-domain Correlation-aware Feedback-free MIMO Transmission and Resource Allocation (CaFTRA) framework tailored for fully-decoupled radio access networks (FD-RAN) to meet the emerging requirements of AI-Native 6G and beyond.
By leveraging artificial intelligence (AI), CaFTRA eliminates the need for real-time channel state information (CSI) feedback during online MIMO operation by predicting CSI from user geolocation.
We introduce a learnable-queries-driven Transformer network for geolocation-to-CSI mapping, which utilizes multi-head attention and learnable query embeddings to accurately capture frequency-domain correlations among resource blocks (RBs), thereby significantly improving CSI prediction precision.
Once base stations (BSs) adopt feedback-free transmission, their downlink transmission coverage can be significantly expanded due to the elimination of frequent uplink feedback.
To fully explore the possibility of such extensive coverage with multi-BS cooperation, we apply a low-complexity many-to-one matching theory-based algorithm, which is proved to converge to a stable matching within limited iterations.
Simulation results based on the Vienna 5G system-level simulator show that CaFTRA achieves stable matching convergence and significant gains in spectral efficiency and user fairness over representative 5G baselines, while remaining effective across different deployment scenario and carrier band through scenario-specific retraining.
\end{abstract}
\IEEEpeerreviewmaketitle

\begin{IEEEkeywords}
6G fully-decoupled radio access network, learnable query embeddings, Transformer-based CSI prediction, feedback-free MIMO transmission, multi-dimensional resource allocation.
\end{IEEEkeywords}

\section{INTRODUCTION}

\IEEEPARstart{F}rom 1G to 5G, each generation of mobile networks has required new spectrum resources, progressively shifting spectrum allocation toward higher frequency bands. This migration inevitably increases path loss, resulting in smaller coverage areas and significantly higher power consumption \cite{Chair-4,Chair-1,6G-1,6G}.
For instance, the coverage area of 5G base station (BS) is only approximately  one-third to one-fourth that of 4G BS, while its power consumption is three to four times higher.
Furthermore, despite the significant power disparity between BSs and mobile devices, the uplink (UL) and downlink (DL) transmissions remain coupled in current cellular architectures, limiting BS coverage primarily by the UL transmission capabilities of mobile devices.
Additionally, current spectrum utilization schemes, such as time division duplex (TDD) and frequency division duplex (FDD), also present inherent limitations \cite{Shen-1,6G-3}.
The TDD bands, while more flexible, have their own limitations, including switching intervals for UL/DL transitions and additional delays introduced by the need to wait for UL/DL time slots.
The FDD bands lack flexibility and efficiency, as fixed portions of the spectrum are designated exclusively for UL or DL use.
Moreover, in cellular networks, factors such as channel state information (CSI) feedback delay and pilot contamination caused by the limited length of reference signal introduce error in channel information estimation \cite{Chair-3,CSI-1,CSI-2,CSI-3,cell-free-book}.
Such error is particularly pronounced in high-mobility scenarios with rapidly varying channels, ultimately leading to performance degradation at the physical (PHY) layer.

The fully-decoupled radio access network (FD-RAN), first proposed in 2019 \cite{JCIN}, has been widely recognized as a key architectural candidate for 6G \cite{chenshixiong}.
In FD-RAN, BSs are physically decoupled into: (1) Control BSs for control services, (2) Uplink BSs for uplink data services, and (3) Downlink BSs for downlink data services. Any spectrum can be used for UL/DL transmission, eliminating the need for FDD guard bands and TDD switching time slots, thereby improving the spectral efficiency (SE).
Through UL/DL decoupling, the coverage area of DL BSs increases dramatically \cite{Dep}, which can reduce operator's cost by deploying less BSs \cite{Cost}.
Through collaboration of multiple BSs, multi-dimensional resources can be coordinated for personalized services with higher network SE.
However, conventional feedback-based transmission methods become infeasible due to this decoupling, necessitating new PHY-layer transmission design.
At the medium access control (MAC) layer, the decoupling of UL/DL functions can dramatically expand the DL BS coverage, enabling flexible multi-BS association and extensive multi-RB resource allocation.
On one hand, the expanded DL BS coverage enables a better strategy of multi-dimensional resource allocation in MAC layer. On the other hand, it increases the complexity of resource allocation algorithm, while the timely acquisition of CSI also remains a challenge.
Moreover, the capability for each user to be simultaneously served by multi-BSs, needs us to redesign MAC-layer resource allocation algorithms.
The key scientific question is how to achieve feedback-free multiple-input and multiple-output (MIMO) transmission and extensive-coverage multi-dimensional resource allocation for FD-RAN.

Motivated by these challenges, in this paper, we develop a frequency-domain Correlation-aware Feedback-free MIMO Transmission and Resource Allocation (CaFTRA) framework specifically designed for FD-RAN.
First, at the PHY layer, inspired by the CSI feedback mechanism of the 3rd Generation Partnership Project (3GPP) New Radio (NR) standards, we propose a Learnable Queries-driven Transformer Network (LQTN) for channel state information (CSI) mapping from user geolocation, leveraging  multi-head attention \cite{TF-1} and learnable query embeddings \cite{TF-2} to capture frequency-domain correlations among resource blocks (RBs), significantly enhancing CSI prediction precision.
Unlike existing 5G networks, the proposed Transformer-based CSI map, based solely on user geolocation, can estimate the CSI parameters of all BSs to UEs for every RB without requiring UE feedback.
Then, at the MAC layer, a many-to-one matching model-based multi-BS association and multi-RB allocation algorithm is proposed for the extented-coverage multi-dimensional resource allocation in FD-RAN.
This algorithm ensures efficient resource scheduling and is analytically shown to converge to stable matching solutions.
The CaFTRA framework enables accurate CSI prediction and outperforms feedback-based transmission schemes in high-mobility scenarios. Furthermore, by leveraging multi-BS cooperation and dynamic RB allocation, CaFTRA substantially enhances network spectral efficiency.

We highlight the main contributions of this work as follows.
\begin{itemize}
\item Under the 3GPP-compatible CSI reporting framework, we design a learnable-queries-driven Transformer network for geolocation-to-CSI mapping, which employs multi-head attention and learnable query embeddings to explicitly capture frequency-domain correlation among RBs. This design enables each BS to jointly predict CSI parameters across all RBs from UE geolocation, without requiring real-time CSI feedback during transmission.

\item Building on the 5G closed-loop spatial multiplexing (CLSM) mode, we propose a geolocation-driven feedback-free MIMO transmission mechanism for DL BSs in FD-RAN, where CSI is predicted from UE geolocation rather than acquired through real-time feedback.

\item To fully explore the possibility of such extensive coverage with multi-BS cooperation, we further develop a practical many-to-one matching-based algorithm for multi-BS association and multi-RB allocation. The proposed algorithm is a low-complexity heuristic with polynomial complexity and finite-iteration convergence to a pairwise-stable matching outcome.

\item We conduct extensive simulations comparing 5G networks and FD-RAN using the well-known Vienna 5G system-level simulator \cite{Vienna5GSLS}. Results show that the proposed CaFTRA enables accurate CSI prediction and outperforms feedback-based schemes in high-mobility cases. It also achieves significant gains in spectral efficiency and user fairness over representative 5G baselines, while remaining effective across different deployment scenario and carrier band through scenario-specific retraining.
\end{itemize}

The remainder of this paper is structured as follows. Section II reviews related works. The system model is introduced in Section III. Section IV details the frequency-domain correlation-aware feedback-free MIMO transmission mechanism. Section V presents the matching-based resource allocation algorithm. Simulation results are provided in Section VI, followed by conclusions in Section VII.

\section{Related Works}

Recently, many efforts have been made to cope with the heavy feedback overheads in massive MIMO systems of 5G.
First, the data-driven CSI feedback compression has gained prominence.
Guo et al. \cite{CSI-0} provided an overview of deep learning-based CSI feedback techniques, highlighting how they reduce feedback overhead through compression and reconstruction.
Nie et al. \cite{CSI-3} proposed a deep learning-based near-field beam training method for extremely large-scale array systems, which efficiently reduced beam training by optimizing the beamformer using pre-estimated CSI without relying on predefined beam codebooks.
Fan et al. \cite{CSI-1} proposed a neural network that disentangled dual-polarized CSI into three components to reduce redundancy and enhance CSI compression and recovery.
Yi et al. \cite{CSI-2} proposed a deep learning-based feedback algorithm for dynamic distributed uplink beamforming in 6G Internet of Vehicles.
This concept has progressed rapidly, with numerous refinements (e.g. attention mechanisms, lightweight models) and even consideration in standards, i.e., 3GPP's Release 18 included a study item on AI-enhanced CSI feedback compression \cite{Chair-2,3GPP-CSI}.

In parallel, CSI prediction has emerged as an important technique to combat feedback latency and outdated channel information.
Here the goal is to forecast future CSI from past observations (e.g. using recurrent or convolutional neural networks), so that the transmitter can proactively obtain CSI without waiting for feedback every time. This approach is also attracting both academic and industrial interest.
Zhou et al. \cite{TF-CSI} proposed a Transformer network-based channel prediction
to forecast future CSI from past observations.
This paradigm shift is not limited to academia, but it is also reflected in industry roadmaps.
For example, Samsung and KDDI recently announced a partnership to integrate AI into distributed MIMO (D-MIMO) for 6G networks, with the aim of enabling self-optimizing, highly adaptive multi-cell MIMO operations.
In the industry, Navabi et al. \cite{CSI-4} investigated the viability of AI techniques for estimating user-channel features (i.e., angle-of-arrival) at a large-array BS, demonstrating the potential of data-driven methods to predict unobserved channel characteristics from observed ones.
Nagao et al. \cite{CSI-5} proposed a technique to estimate path loss by extracting features from map images around the receiver, using the Hough transform to calculate road angles and widths.
Likewise, industry alliances such as the O-RAN Alliance (with its AI/ML RAN focus) and the Next G Alliance are actively promoting AI-native network design, from the PHY layer up through RAN control.
Different from existing works, we investigate a data-driven MIMO transmission mechanism that requires no channel feedback, determining all MIMO transmission parameters for any user's geolocation, rather than focusing on partial channel features.

Since FD-RAN was first introduced in \cite{JCIN} for 6G, some research efforts have emerged.
Yu et al. \cite{FD-1} proposed a two-stage DL channel estimation method and a dynamic resource cooperation framework for FD-RAN, leveraging multi-connectivity and coordinated beamforming to maximize the weighted sum achievable rate.
Qian et al. \cite{FD-2} proposed a joint UL/DL resource scheduling scheme for FD-RAN, integrating dynamic spectrum division, user-BS-subchannel matching, and power control to optimize UL and DL asymmetric service.
Xu et al. \cite{FD-3} proposed a joint multiple access collaboration and power management solution over FD-RAN to accelerate federated learning (FL) in end-cloud two-tier computing, optimizing UL and DL transmission through multi-BS access and power control, significantly improving FL training efficiency in wireless networks.
However, these work primarily assume that the DL BSs can obtain complete channel information through the control BS, which is impractical in real-world MIMO transmissions due to the large scale of BS antennas and delay.
Recently, Liu et al. \cite{FD-0} proposed an end-to-end data-driven MIMO solution for FD-RAN, eliminating conventional channel feedback by mapping geolocation to MIMO transmission parameters through codebook-based and non-codebook-based approaches, with assumption of historical complete channel.
Xu et al. \cite{FD-4} proposed a feedback-free coordinated multi-BS transmission framework for FD-RAN, leveraging hierarchical reinforcement learning, transformer-based subband processing, and diffusion modeling to optimize MIMO parameters using only user's geolocations.
However, these works assume that the downlink BSs can obtain historical complete downlink channel information for UEs at certain geolocations. Moreover, they do not explicitly exploit the frequency-domain correlation among RBs, since the CSI-related inference is performed at the RB/subband level rather than through joint cross-RB prediction. In contrast, the main novelty of this work lies in the joint design of: (i) a frequency-domain correlation-aware geolocation-to-CSI prediction framework, in which the proposed LQTN jointly models CSI across all RBs through learnable query embeddings; and (ii) a geolocation-driven matching-based multi-BS and multi-RB resource-allocation mechanism for FD-RAN.

\section{System Model}

\begin{figure}[H]
\centering
\includegraphics [width=3.3in]{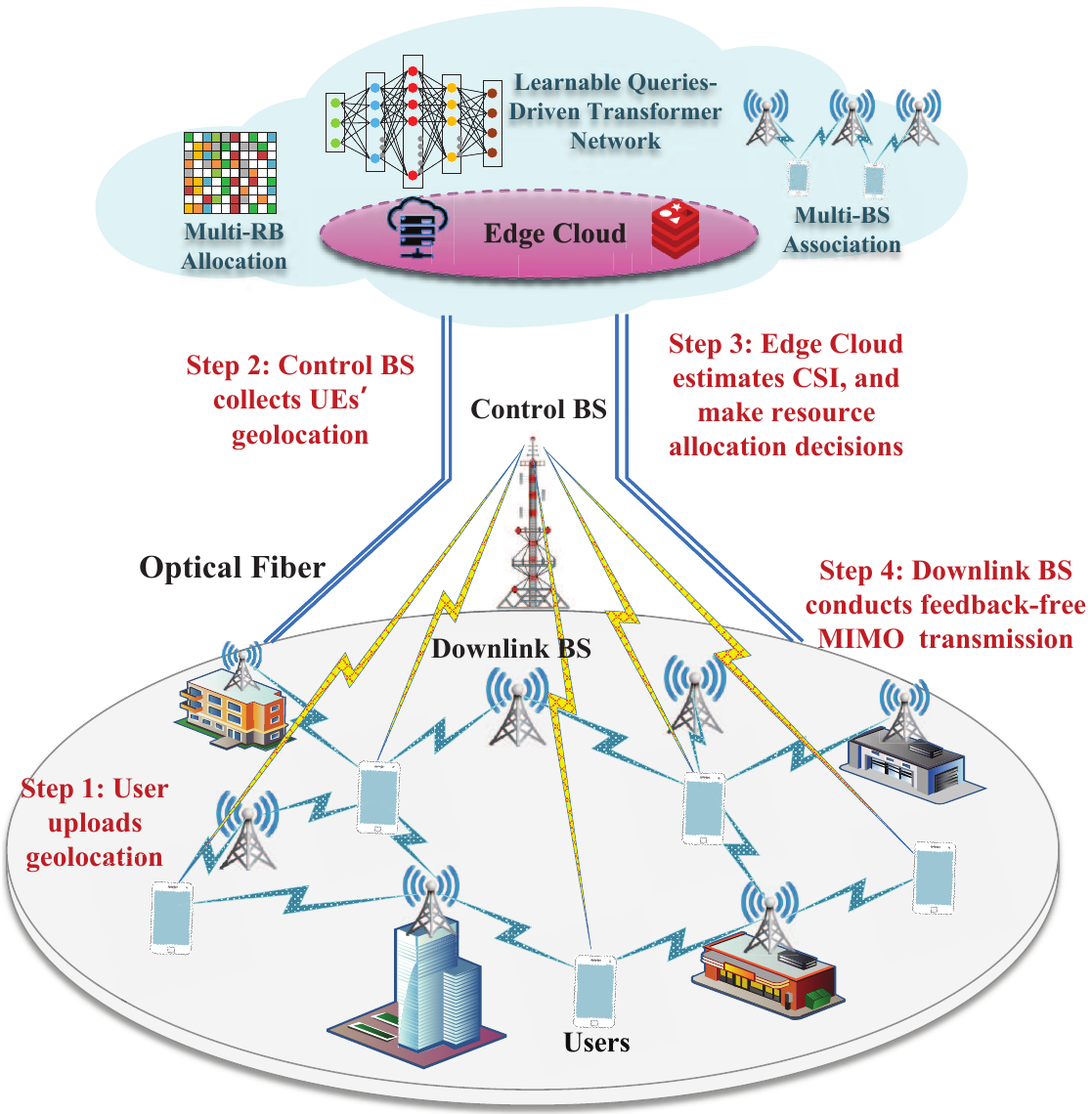}
\centering
\caption{The Proposed CaFTRA Framework in FD-RAN.}
\label{net}
\end{figure}

Fig. \ref{net} depicts the CaFTRA framework within an FD-RAN scenario, encompassing learning-oriented feedback-free MIMO transmission, joint multi-BS association and multi-RB allocation for extended DL coverage.
This architecture physically decouples the UL and DL functionalities across different BSs to optimize resource utilization and enhance overall network flexibility.
For clarity, UL BSs are omitted in the figure, as this work focuses exclusively on the DL transmission and resource scheduling of FD-RAN.

\begin{figure}[H]
\centering
\includegraphics [width=3.5in]{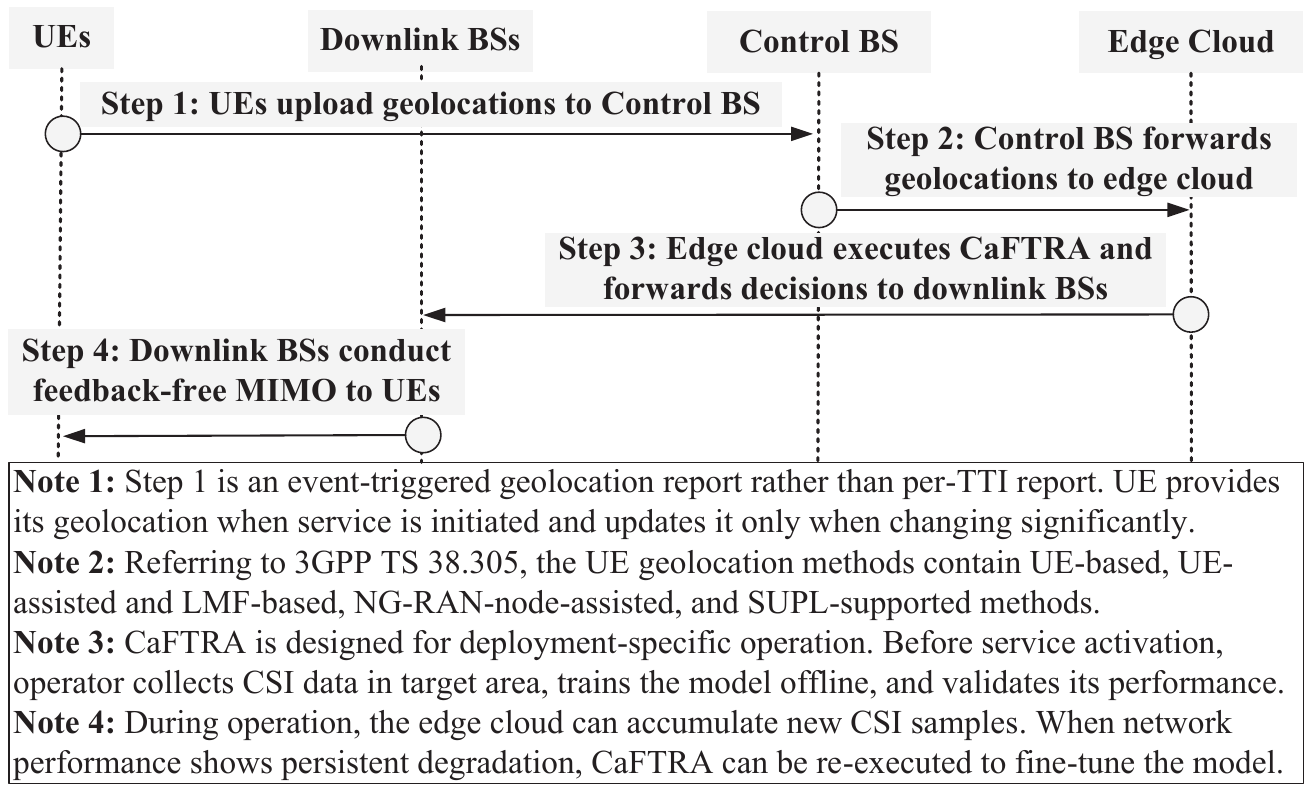}
\centering
\caption{End-to-end Control-plane Signaling Procedure of CaFTRA.}
\label{R1-Control}
\end{figure}

To make the signaling interpretation more explicit, Fig. \ref{R1-Control} further illustrates the end-to-end control-plane workflow considered in this work.
This also highlights the practical relevance of the proposed framework under the AI-native RAN paradigm \cite{AIRAN,AIRAN2}, where edge-cloud-based intelligence is expected to play a key role in enabling scalable and low-latency RAN optimization.

\subsection{Feedback-based MIMO Transmission Mechanism in 3GPP}

We first introduce the signal process in the orthogonal frequency-division multiplexing (OFDM) defined in 3GPP TS 38.214 \cite{3GPPTS38.214}.
At sampling time instant $t$, the received symbol vector $\mathbf{y}_{k,t}\in \mathbb{C}^{N_R\times 1}$ on subcarrier $k$ is given by:
\begin{equation}
\mathbf{y}_{k,t}=\mathbf{H}_{k,t} \mathbf{W}_{i} \mathbf{x}_{k,t}+ \mathbf{n}_{k,t},~k=1,\cdots, K, ~t=1,\cdots,T,
\end{equation}
where $\mathbf{H}_{k,t}\in \mathbb{C}^{N_R\times N_T}$ denotes the channel matrix on subcarrier $k$ at time instant $t$, $\mathbf{W}_{i}\in \mathcal{W}$ is the precoding matrix with  $i$ denoting the index within the codebook of precoding matrices $\mathcal{W}$, $\mathbf{x}_{k,t}\in \mathcal{A}^{L \times 1}$ is the transmit symbol vector with $\mathcal{A}$ being the utilized symbol alphabet, and $\mathbf{n}_{k,t}\in \mathcal{CN}(0,\sigma_n^2\mathbf{I})$ is the white complex-valued Gaussian noise with variance $\sigma_n^2$, $\mathbf{I}\in \mathbb{R}^{N_R\times N_R}$ is the identity matrix.
The dimension of the transmit symbol vector depends on the number of useful spatial transmission layers $R$.

Then, the received symbol vector $\mathbf{y}_{k,t}$ will be filtered by a linear equalizer given by a matrix $\mathbf{F}_{k,t}\in \mathbb{C}^{L \times N_R}$.
The channel equalization $\mathbf{F}_{k,t}$ could recover the received signal to the original transmitted signal.
The linear receiver is typically chosen according to a zero forcing (ZF) or minimum mean square error (MMSE) design criterion. The input signal vector is normalized to unit power.
In this paper, we consider the zero-forcing equalizer, which leverages the pseudo inverse of effective channel matrix as follows:
\begin{equation}
\mathbf{F}_{k,t}(\mathbf{W}_{i}) = \left[ \left( \mathbf{H}_{k,t} \mathbf{W}_i \right)^{\mathrm{H}} \mathbf{H}_{k,t} \mathbf{W}_i \right]^{-1} \left( \mathbf{H}_{k,t} \mathbf{W}_i \right)^{\mathrm{H}},
\end{equation}
where $(\cdot)^{\mathrm{H}}$ represents the Hermitian transpose operation.

Subsequently, the equalized output of this filter is the post-equalization symbol vector $\mathbf{p}_{k,t}$:
\begin{equation}
\mathbf{p}_{k,t} = \mathbf{F}_{k,t} \mathbf{y}_{k,t} = \underbrace{\mathbf{F}_{k,t} \mathbf{H}_{k,t} \mathbf{W}_i }_{\mathbf{G}_{k,t} \in \mathbb{C}^{L \times L}} \mathbf{x}_{k,t} + \mathbf{F}_{k,t} \mathbf{n}_{k,t},
\end{equation}
where $\mathbf{G}_{k,t}(\mathbf{W}_{i})\triangleq\mathbf{F}_{k,t} \mathbf{H}_{k,t} \mathbf{W}_{i}$ could recover the received signal for $L$ spatial transmission layers.

In this way, the post-equalization SINR on layer $l$ is expressed as:
\begin{equation}\label{SINR}
\text{SINR}_{k,t,l}(\mathbf{W}_{i}) = \frac{\left| \mathbf{G}_{k,t}(l,l) \right|^2}{\sum_{i \neq l} \left| \mathbf{G}_{k,t}(l,i) \right|^2 + \sigma_n^2 \sum_i \mathbf{F}_{k,t}(l,i)},
\end{equation}
where $\mathbf{G}_{k,t}(l,i)$ refers to the element in the $l$-th row and $i$-th column of matrix $\mathbf{G}_{k,t}\in\mathbb{C}^{L \times L}$.
The first term in the denominator corresponds to inter-stream interference, and the second term accounts for noise enhancement.

\subsection{CSI Parameter Design for MIMO Transmission}

According to the 3GPP TS 38.214 \cite{3GPPTS38.214}, reliable MIMO transmission depends on accurate CSI to determine transmission parameters, which includes:
\begin{enumerate}
  \item \textbf{Rank Indicator (RI)}: Indicates the number of spatial data streams supported by current channel conditions.
  \item \textbf{Channel Quality Indicator (CQI)}: Suggests the appropriate channel coding rate and modulation scheme.
    \item \textbf{Precoding Matrix Indicator (PMI)}: Identifies the optimal precoding matrix index from a predefined codebook.
\end{enumerate}

The selection of these CSI parameters is typically a sequential process.
First, for each RB, the optimal precoding matrix is determined by maximizing the mutual information derived from the post-equalization SINR (see Eq.~\eqref{SINR}), as established by Shannon's theory. The RI and PMI selection depend on the employed codebook, with Type-I codebooks being the most widely adopted in 3GPP standards (details will be introduced later). Once RI and PMI are determined, the CQI is chosen to ensure that the block error rate (BLER) does not exceed 10\%.

Referring to Eq. \eqref{SINR} and well-known Shannon Theory \cite{Shan}, we can calculate the post-equalization mutual information in terms of the post-equalization $\text{SINR}_{k,t,l}$ as
\begin{equation}
I_{k,t}(\mathbf{W}_{i}) = \sum_{l=1}^{L} \log_2 \left[ 1 + \text{SINR}_{k,t,l}(\mathbf{W}_{i}) \right]
\end{equation}

The optimal precoding matrix $(\mathbf{W}_{i})$ is selected by maximizing the mutual information over all consider RBs, i.e., over spectrum range subcarrier-$k\in\{1,\cdots, K\}$ and temporal-range time slot-$t\in\{1,\cdots, T\}$:
\begin{equation}\label{PMI_CAL}
\begin{split}
&\max\limits_{\mathbf{W}_i}~~~~~~~~~\sum_{k=1}^{K} \sum_{t=1}^{T} I_{k,t}(\mathbf{W}_i)\\
&~~~\textrm{s.t.}
~~~\mathbf{W}_i \in \mathcal{W},~i\in \{1,2,...,|\mathcal{W}|\},
\end{split}
\end{equation}
where $\mathcal{W}$ is the pre-designed codebook, and the optimal solution $\mathbf{W}^*_j$ of problem \eqref{PMI_CAL} needs to be selected by exhaustive search within it.

Referring to 3GPP NR standard, the UE computes the post-equalization mutual information for all possible precoders (consists of RI and PMI) from the pre-designed codebook.
The optimal precoding matrix $\mathbf{W}^*_j$ is chose to maximizing the sum mutual information over the RB, where the RI is given by this layer number and the PMI is the indice within the codebook.

For the selection of CQI, the UE first averages the post-equalization SINR across the relevant frequency-time resources within the RB.
The Effective SINR Mapping (ESM) methods is then employed to map the set of SINR values to an equivalent SNR for a single-input single-output (SISO) AWGN channel, ensuring comparable block error rate performance to the original OFDM system \cite{SNR}.
The ESM can be formulated as follows:
\begin{equation}\label{ESM}
\text{SNR}_{\text{ESM}} = \gamma f^{-1} \left( \frac{1}{KTL} \sum_{k=1}^{K}\sum_{t=1}^{T}\sum_{l=1}^{L} f \left( \frac{\text{SINR}_{k,t,l}}{\gamma} \right) \right),
\end{equation}
where mapping function $f$ is the bit interleaved coded modulation (BICM) capacity in the well-known mutual information effective SNR mapping, and the CQI dependent $\gamma$ value is the calibration factor that adjusts the mapping to the different code rates and modulation alphabets \cite{ff}.

Finally, according to 3GPP TS 38.214 \cite{3GPPTS38.214}, each CQI corresponds to a pre-designed modulation and coding scheme (MCS).
The CQI feedback value is the highest possible value (ranging from 0 to 15) with block error rate is no more than 10\% for the equivalent SISO AWGN channel \eqref{ESM}.
It is worth noting that, in the 3GPP standard, the downlink MIMO transmission has two codewords (CWs), i.e., CW0 is used by every channel, and CW1 is used by the user data when spatial multiplexing is enabled.
Therefore, there are two CQI parameters corresponding to CW0 and CW1, denoted as CQI 1 and CQI 2 in this paper.

\subsection{Codebook Design for PMI}

To facilitate the design of a neural network for CSI prediction, it is essential to provide a detailed explanation of parameters that constitute the PMI. In this paper, we focus on the Type I codebook, the most common MIMO codebook in 3GPP NR, which inherits key design principles from the LTE codebook.
For completeness, we note that other codebooks in the 3GPP NR standard, such as Type II and Enhanced Type II, are primarily intended for multi-user MIMO (MU-MIMO) scenarios and can be incorporated in a similar manner.
Extending the framework to MU-MIMO requires additional user pairing and interference-aware scheduling, which is left for future work.
Since PMI remains a discrete codebook index, the proposed geolocation-to-CSI prediction formulation naturally remains applicable when the codebook dimension changes, although a larger codebook generally requires more training data to maintain high prediction accuracy.

The type I codebook employs a two-stage structure, $W= W_1 \times W_2$, and the codebook is based on beam selection:
\begin{itemize}
  \item $W_1$: Selects a wideband beam group based on the long-term, wideband spatial characteristics of the channel.
  \item $W_2$: Selects beams based on the short-term subband characteristics of the channel and quantifies the phase difference between dual-polarization directions to achieve in-phase combination between polarization directions.
\end{itemize}

Firstly, the UE needs to determine the beam set.
The spatial dimension's orthogonal basis is composed of $N_1$  discrete Fourier transform (DFT) beams, each of length $N_1$, refined by oversampling with a rotation factor $R(q_1)$ at an oversampling rate of $O_1$. Similarly, the second spatial dimension's orthogonal basis is composed of $N_2$ DFT beams, each of length $N_2$, refined by oversampling with a rotation factor $R(q_2)$ at an oversampling rate of $O_2$.

Within the Type I codebook, the UE is permitted to report back one beam $L=1$ out of the available grid of beamformings (GoBs) for the given configuration \cite{Code}. According to 3GPP NR standard, the UE reports back to the BS with the help of the four indices:
\begin{itemize}
  \item $i_{1,1}$: Gives information about index of the beam in azimuth dimension.
  \item $i_{1,2}$: Gives information about index of the beam in elevation dimension.
  \item $i_{1,3}$: For 2,3,4 layers, this gives information on designing the layers with 2,3 or 4 one layered beam.
  \item $i_{2}$: Controls the co-phasing between the polarization's at the subband level. This information helps to adapt according to the channel variations.
\end{itemize}

These four indices are subsequently mapped to the precoding matrix computation to select the optimal precoder.

\section{Frequency-Domain Correlation-Aware Feedback-Free MIMO Transmission Mechanism}

In this section, we first introduce the proposed feedback-free MIMO transmission mechanism. Next, we present the construction of the Transformer-based CSI map, which leverages multi-head attention and learnable query embeddings to accurately capture frequency-domain correlations among RBs.

\subsection{Geolocation-Driven Feedback-Free MIMO Transmission}

\begin{figure}[H]
\centering
\includegraphics [width=3.5in]{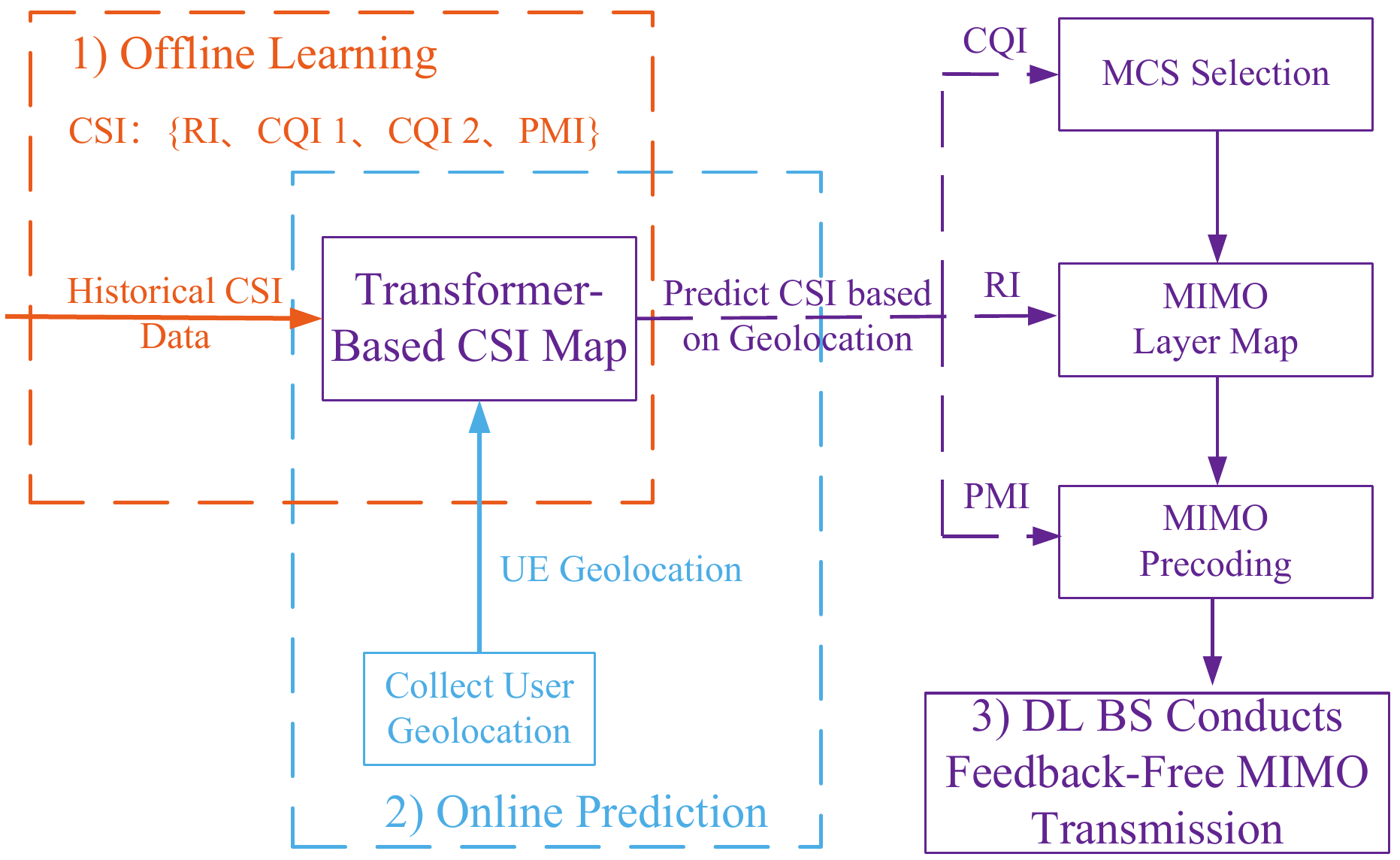}
\centering
\caption{Workflow of the Proposed Geolocation-Driven Feedback-Free MIMO Transmission Mechanism.}
\label{workflow}
\end{figure}

Since the CSI parameters are all discrete integers, CSI prediction can be naturally  transformed into a multi-objective classification problem.
As illustrated  in Fig. \ref{workflow}, we develop a Transformer-based CSI map model that enables feedback-free MIMO transmission by exploiting the powerful modeling capabilities of Transformer networks.

After training the Transformer-based CSI map using historical labeled data, the BS geolocation $\mathbf{BS}_{loc}$ and the available UE geolocation $\mathbf{UE}_{loc}$ are used as the inputs to the network.
The encoder and decoder modules then produce the predicted CSI parameters for all RBs, i.e., $\mathrm{CSI}_i=[\mathrm{RI}_i,\mathrm{CQI1}_i,\mathrm{CQI2}_i,\mathrm{PMI}_i]^T$, $i=1,2,\cdots,RB_{num}$. Historical CSI samples are collected asynchronously for offline model construction and maintenance, rather than for per-transmission online adaptation.

The overall feedback-free MIMO transmission process for DL BSs consists of the following steps:
\begin{enumerate}
\item \textbf{Offline construction of the Transformer-based CSI map:} For each DL BS, an independent Transformer network is constructed and trained offline using historical geolocation-to-CSI labeled data, serving as the CSI predictor for that deployment area.

\item \textbf{Online CSI prediction based on UE geolocation:} During operation, the edge cloud obtains UE geolocation information using standard UE positioning methods defined in 3GPP TS 38.305, predicts CSI, makes resource-allocation decisions, and forwards the corresponding decisions to the DL BSs.

\item \textbf{Feedback-free MIMO transmission using predicted CSI:} With the predicted CSI parameters, and following the 5G CLSM procedure, each DL BS conducts MIMO transmission without requiring real-time CSI feedback from the UE.
\end{enumerate}

\subsection{Construction of Learnable Queries-Driven Transformer Network}

The success of large language models has already demonstrated the Transformer's powerful capability in capturing complex correlations, which motivates us to employ it for modeling both spatial and frequency-domain dependencies in CSI prediction.
To predict the CSl across multiple RBs in a dynamic wireless communication environments, we propose a Learnable Queries-driven Transformer Network (LQTN) that leverages spatial information between the BS and UE. The model is designed to capture both the spatial correlation among UEs and the frequency-domain correlation of the RBs.

\begin{figure}[H]
\centering
\includegraphics [width=3.5in]{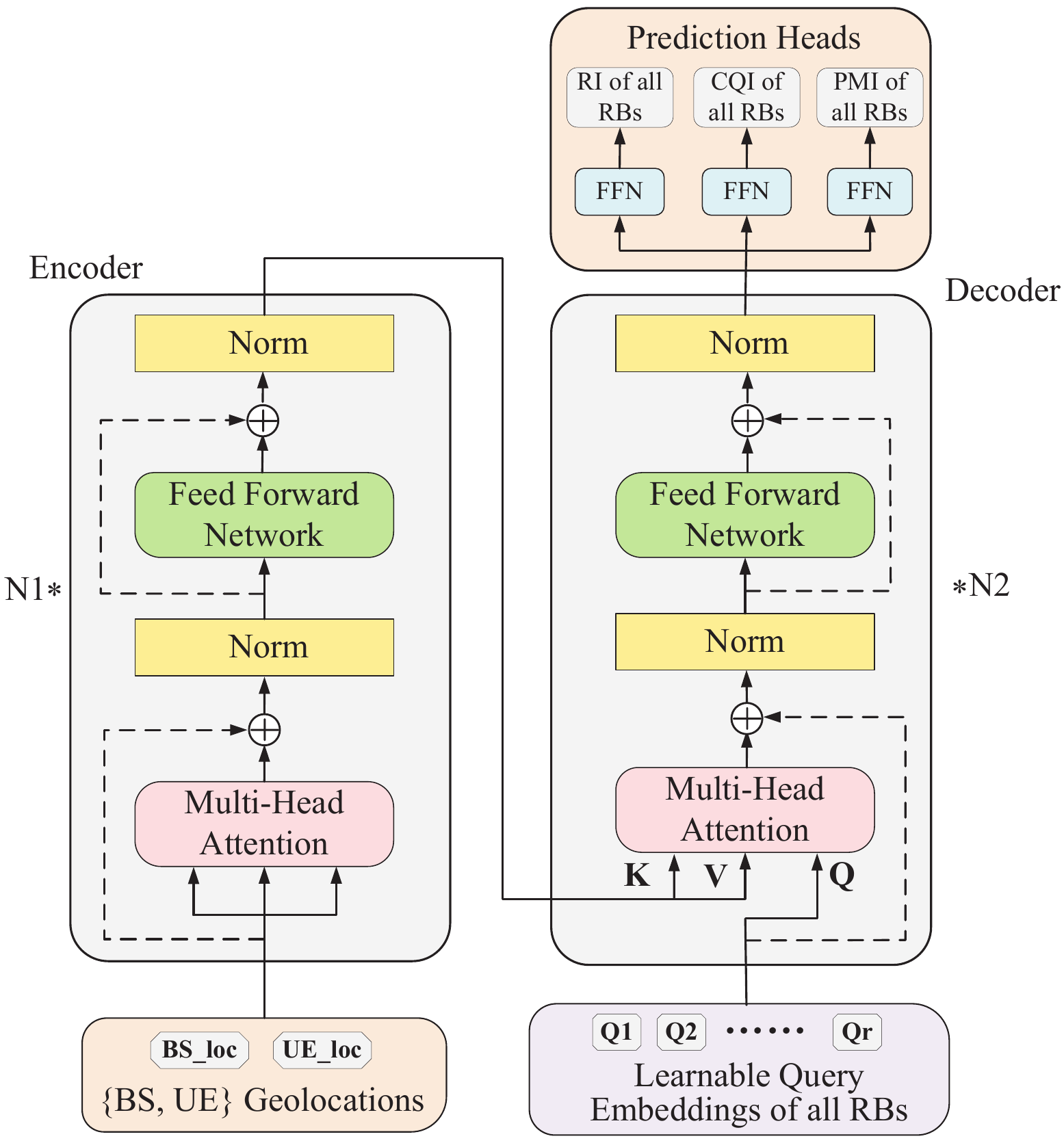}
\centering
\caption{Schematic Diagram of the LQTN-based CSI Prediction Model.}
\label{Transformer}
\end{figure}

\textbf{Overall Network Structure}:
The overall architecture follows an encoder-decoder design tailored for multi-RB CSI prediction. Fig. \ref{Transformer} provides a schematic diagram of the LQTN-based CSI prediction model structure, which consists of three main components: BS-UE Position Encoder, RB-Aware decoder and CSI Prediction Head. Given the spatial positions of the BS and UEs as input (i.e., $(BS_{loc})$ and $(UE_{loc})$), the model first extracts high-level spatial features through an encoder, then uses a set of RB-specific queries in the decoder to obtain per-RB latent representations, which are finally processed by individual prediction heads to produce CSI parameters for each RB.
The core of both encoder and decoder is the multi-head attention mechanism. 

\textbf{BS-UE Position Encoder}:
The encoder is responsible for extracting spatial representations from the input positions of the BS and UEs. Each location vector is first projected into a high-dimensional feature space using a shared feed-forward network.
This is followed by a multi-head attention module that models the spatial relationships among all nodes, enabling context-aware feature learning. The output of the encoder is a set of rich, context-aware embeddings for each UE. 	

At the core of both the encoder and decoder blocks lies the multi-head attention mechanism, which is instrumental in enabling the model to capture global contextual information and complex dependencies across inputs. In the encoder, the multi-head attention layer computes attention scores among all position features, thereby modeling interactions between every pair of BS and UE locations. This not only helps in capturing the inter-geolocation dependencies but also enhances the ability to learn spatial correlation patterns crucial for accurate CSI estimation.

\textbf{Frequency-Domain Correlation-Aware Decoder}:
To model the CSI characteristics across multiple RBs, we employ a learnable query based decoder. Each RB is associated with a set of learnable embeddings that acts as a vector in the decoder's query. These queries attend to the encoder's outputs (used as keys and values), allowing each RB representation to selectively integrate spatial context relevant to CSI prediction. The decoder outputs RB-specific latent vectors, which are further processed by individual prediction heads to estimate the desired CSI parameters for each RB.

In the decoder, the multi-head attention module is even more critical. By attending to both its own queries and the encoded geolocation representations, the decoder can dynamically weigh and aggregate information, thereby extracting frequency-domain correlation and interactions unique to each RB. The use of multiple attention heads allows the model to attend to information from different subspaces and perspectives simultaneously, providing a rich and nuanced understanding of both spatial and frequency correlations necessary for high-fidelity CSI prediction.

A key innovation in the decoder is the introduction of learnable query embeddings \cite{TF-2}, where each vector in the query $(Q_1, Q_2, \cdots Q_r)$ corresponds to a specific RB, and the entire query sequence act as the input of decoder.
Unlike traditional approaches that use fixed positional encodings or static queries, these queries are trainable parameters that are optimized jointly with the rest of the network. During inference, the learnable queries serve as dynamic pointers, each focusing attention on the prediction for its respective RB.

\textbf{CSI Prediction Heads}:
The RB representations are then passed to three CSI prediction heads. Each head consists of several Feed Forward Network (FFN) layers and corresponds to a particular aspect of the CSI, i.e., RI, CQI, and PMI. The CSI prediction heads include:
\begin{itemize}
  \item \textit{RI Prediction Head}: Estimates the RI across all RBs, which reflects the supported number of spatial streams. Accurate RI prediction is essential for determining the spatial multiplexing capabilities in each RB.
  \item \textit{CQI Prediction Head}: Provides a fine-grained assessment of channel quality that informs adaptive modulation and coding strategies. High-quality CQI prediction leads to improved link adaptation and spectral efficiency.
  \item \textit{PMI Prediction Head}: Predicts the optimal precoding matrices for each RBs, facilitating efficient spatial beamforming and maximizing system throughput.
\end{itemize}

Overall, this architecture jointly captures spatial relationships and RB-wise spectral dependencies, enabling accurate and fine-grained CSI prediction across frequency resources.

\section{Extensive-Coverage Multi-Dimensional Resource Allocation}

To assess the capacity gains of FD-RAN over conventional 5G networks, we investigate a fundamental question: under the condition of ensuring the minimum QoS (i.e., each UE is allocated at least $Q$ RBs), we compare the network capacity of FD-RAN and 5G network using the same spectrum resources.

In practical systems, the PMI selected by the transmitter is often imperfect due to limited channel knowledge, estimation errors, or environmental variations. This inaccuracy can lead to packet losses, resulting in actual throughput being lower than the Shannon capacity. Therefore, to accurately assess system-level performance, we adopt a simulation-based approach (i.e., the Vienna 5G system-level simulator) to calculate throughput, rather than relying solely on the Shannon capacity.

\subsection{Estimation of Maximum PHY-Layer Transmission Rate Based on CSI}

We first establish  a model for estimating the maximum PHY-layer transmission rate based on CSI parameters, which provides a unified performance metric for resource allocation in the proposed CaFTRA framework.

Table 5.2.2.1-2 in 3GPP TS 38.214 \cite{3GPPTS38.214} specifies the modulation schemes and code rates corresponding to different CQI values.
Referring to the PHY-layer frame structure of OFDM, we can give the calculation procedure for estimating the maximum PHY-layer transmission rate  based on CSI parameters.
For example, when $CQI = 1, RI=1$ for a given RB, the modulation scheme is QPSK with a code rate of 0.076 according to Table 5.2.2.1-2 in \cite{3GPPTS38.214}. The calculation procedure for the maximum PHY-layer transmission rate is:
\begin{enumerate}
  \item
\textbf{Number of Resource Elements (RE) per RB}: (14 OFDM symbols per subframe) $\times$ (1 RB $\times$ 12 REs per symbol) $=168$ REs per subframe.
  \item
\textbf{Adjusting for Physical Downlink Control Channel (PDCCH)}:
Since there are 3 PDCCH symbols per subframe, the number of REs occupied by the Control Format Indicator (CFI) needs to be subtracted: 168 REs per subframe $-$ (3 PDCCH symbols $\times$ (1 RB  $\times$  12 REs per symbol)) $=132$ REs per subframe.
  \item
\textbf{Bits per Subframe Based on Modulation}:
The modulation order for QPSK is 2, i.e., $2 \times 132 = 264$ bits per subframe.
  \item
\textbf{Transmission Block Size (TBS) Based on Code Rate}:
The number of information bits in transmission block is:
TBS $=$ Total bits in the physical channel $\times$ code rate $= 264 \times 0.076 = 20.064$ bits.
  \item
\textbf{Maximum PHY-layer Transmission Rate }:
For FDD frame structure, the downlink peak rate is calculated as:
20.064 (TBS) $\times$ 1 (number of streams) $\times$ 10 (downlink slots) $\times$ 100 (frames per second) $= 20064$ bit/s $= 0.020064$ Mbps.
\end{enumerate}

The maximum PHY-layer transmission rate for other CQI and RI values can be calculated using the same procedure.
Provided that the PMI is accurately predicted, the frame error rate remains low, making the estimated rate practically achievable.
Therefore, once the CSI parameters are known, the maximum PHY-layer transmission rate (hereafter referred to as rate) can be obtained and used as the basis for resource allocation at the MAC layer.
We can see that the rate represents the upper bound of throughput under ideal conditions where no transmission errors occur.
The output of the model (i.e., the maximum PHY-layer transmission rate) corresponds to the function $Rate(CSI)$.

\subsection{Multi-BS Association and Multi-RB Allocation Based on Many-to-One Matching Theory}

The objectives of extensive-coverage, multi-dimensional resource allocation in MAC layer can vary widely, especially when catering to personalized user services, resulting in diverse performance metrics. To compare the capacity differences between CaFTRA-based FD-RAN and 5G networks, we focus on a fundamental question: under the condition of ensuring minimum resource (i.e., each UE is allocated at least $Q$ RBs), we aim to compare the network capacity of CaFTRa-based FD-RAN under CaFTRA framework and 5G NR using the same spectrum resources.

Suppose we have $x$ BSs, and each BS have $y$ RBs, then the total number of RBs can be allocated to users is $W= x \times y$, and we can bind BSs and RBs together.
Note here we assume that the whole spectrum band is equally divided to $x$ BSs, and they do not share spectrum with each other in FD-RAN to avoid interference in the considered area.
Thus, there are a total of $W$ BS-RBs and $M$ UEs, and define a binary variable $\mathbf{X}(w,m)$ to indicate whether BS-RB $w$ is selected to serve UE $m$, which is expressed as
\begin{equation}
\mathbf{X}(w,m)=\left\{
\begin{split}
&1, ~~~~\mbox{if BS-RB $w$ serves UE $m$},\\
&0, ~~~~\mbox{otherwise}.
\end{split}
\right.
\end{equation}

The following formulation provides an SE-oriented optimization objective for benchmarking purposes, while the practical scheduler developed in this work is a matching-based heuristic rather than a globally optimal solver.
To maximize the sumrate of network, we model the extensive-coverage multi-dimensional resource allocation, i.e., the three-dimensional allocation problem of multi-BSs, multi-RBs, and multi-UEs as a large-scale 0-1 integer programming problem as follows:
\begin{equation}
\begin{split}
&\max\limits_{\mathbf{X}(w,m)\in \{0,1\}}~~~~\sum_{r=1}^R \sum_{u=1}^U \mbox{Rate}[CSI_{LQTN}(\mathbf{X}(w,m))]\\
&~~~\textrm{s.t.}
~~~~~~~~\sum_{m=1}^M \mathbf{X}(w,m)=1,~\forall w=1,\cdots,W \\
&~~~~~~~~~~~~~~~\sum_{w=1}^W \mathbf{X}(w,m)\geq Q,~\forall m=1,\cdots,M,
\end{split}
\end{equation}
where the objective function $\mbox{Rate}[CSI_{LQTN}(\mathbf{X}(w,m))]$ represents the estimated PHY-layer rate of BS-RB pair $w$ serving UE $m$ using LQTN to obtain CSI, the two constraints ensure that each RB serves only one UE and each UE is served by at least $Q$ RB. This problem is non-convex and NP-hard \cite{Match}. To find the optimal solution of $\mathbf{X}$, we would need to exhaustively search all possible combinations of RBs and BSs scheduling to users. This approach is impractical in real systems due to its large scale and distinct discrete nature. However, since $\mathbf{X}$ is binary variable, we can formulate the allocation of BSs and RBs as a matching problem.
Matching theory, which has been recognized by a Nobel Prize in Economics, provides a mathematically tractable solution to combinatorial matching problems between participants in two distinct sets, using each participant's individual information and preferences.

Since each user can select multiple RBs and BSs, and each RB can serve only one user, we can transform the UE-BS-RB matching problem into a large-scale many-to-one matching problem \cite{TMC-Match}. By binding BSs and RBs into BS-RB pairs and performing many-to-one matching with multiple users, the matching relationships represent the results of multi-BS association and multi-RB allocation. In the following sections, we address the mapping problem of BS-RBs and UEs based on the many-to-one matching model, optimizing $\mathbf{X}$.

\textbf{\emph{Definition 1:}} A mapping $\mu$ from UEs ($\mathcal{M}$) to BS-RB pairs ($\mathcal{W}$) is called a many-to-one matching if, for any $m \in \mathcal{M}$ and $w \in \mathcal{W}$:
\begin{description}
\item[-] \hspace{-6mm}$\mu(m) \subseteq \mathcal{W}$, the set of BS-RB pairs matched to UE $m$;
\item[-] \hspace{-6mm}$\mu(w) \subseteq \mathcal{M}$, the (unique) UE matched to BS-RB $w$;
\item[-] \hspace{-6mm}$m \in \mu(w)$ if and only if $w \in \mu(m)$.
\end{description}

Each UE $m$ can be matched to a subset of BS-RB pairs, and each BS-RB can be matched to at most one UE. Given a subset of potential BS-RB pairs $\hat{\mathcal{W}} \subseteq \mathcal{W}$, user $m$'s choice set is $\mathcal{W}_m(\hat{\mathcal{W}})$. We further define:

\textbf{\emph{Definition 2:}} A matching $\mu$ is pairwise stable if there does not exist any pair $(m,w)$ (with $m\notin \mu(w)$, $w\notin \mu(m)$) such that both would strictly prefer to be matched with each other over their current matches.

\textbf{\emph{Definition 3:}} The preference of a BS-RB $w$ is said to be substitutable if, for any $m, m' \in \mathcal{W}_w(\hat{\mathcal{M}})$, $m$ remains in $w$'s choice set even after removing $m'$ from consideration.

Motivated by these properties, we adopt a practical many-to-one matching-based heuristic, denoted as many-to-one matching model-based multi-BS association and multi-RB allocation algorithm (M3-MAMA) to obtain a pairwise-stable allocation with polynomial complexity and finite-iteration convergence.

\begin{algorithm}[!h]
    \caption{Many-to-one Matching Model-based Multi-BS Association and Multi-RB Allocation (M3-MAMA) }
    \begin{algorithmic}[1]
        \REQUIRE UE set $\mathcal{M}$, BS-RB set $\mathcal{W}$, number of BS-RB pairs $W$.
        \ENSURE Optimized UE-BS-RB allocation strategy $X^*$.
        \STATE \textbf{Initialization Phase:}
        Generates preference lists of UEs by estimating the maximum achievable PHY-layer rate for available BS-RBs, and tentatively applies to up to $Q$ unallocated BS-RBs.
        Each unallocated BS-RB is then assigned to the UE with the highest estimated rate.
        \STATE \textbf{Exchange Matching Phase:}
        \FOR{$i=1,\cdots,W$}
        \FOR{$j=1,\cdots,W$}
        \STATE  Compute the current total system throughput $T_0$.
        \STATE  If BS-RB $i$ and BS-RB $j$ are assigned to the same UE, skip to the next iteration.
        \STATE If BS-RB $i$ and BS-RB $j$ are assigned to different UEs, attempt the following exchanges:
        \STATE  \textbf{Exchange Matching Attempt 1:} Swap the UEs assigned to  BS-RB $i$ and BS-RB $j$, and compute  the new  throughput $T_1$.
        \STATE  \textbf{Exchange Matching Attempt 2:} Assign the UE from BS-RB $j$ to BS-RB $i$. If this assignment satisfies the problem constraints, compute the new throughput as $T_2$; otherwise, set $T_2 = 0$.
        \STATE  \textbf{Exchange Matching Attempt 3:} Assign the UE from BS-RB $i$ to BS-RB $j$. If this assignment satisfies the problem constraints, calculate the new throughput as $T_3$; otherwise, set $T_3 = 0$.
        \STATE  Compare $T_0, T_1, T_2,$ and $T_3$, and select the exchange matching attempt that yields the highest throughput. Update $\mathbf{X}$ and proceed to the next iteration.
        \ENDFOR
        \ENDFOR
    \end{algorithmic}\label{match}
\end{algorithm}

\textbf{\emph{Lemma 1:}}
The M3-MAMA algorithm is guaranteed to converge to a pair-wise stable matching solution.

\textbf{\emph{Proof:}}
Suppose, for contradiction, that there exists a UE $m$ and a BS-RB $w$ such that $m \notin \mu(w)$ and $w \notin \mu(m)$, and $\phi \in \mathcal{W}_m(\mu(m) \cup {w})$, $\phi \in \mathcal{W}_w(\mu(w) \cup {m})$, while ${\phi} \succ_m \mu(m)$ and ${\phi} \succ_w \mu(w)$ hold true.

On one hand, ${w} \succ_m \mu(m)$ means that UE $m$ must have made a matching request to BS-RB $w$ at some iteration. On the other hand, $m \notin \mu(w)$ and $w \notin \mu(m)$ both hold simultaneously. Therefore, when user $m$ made the request, we can conclude that either BS-RB $w$ rejected UE $m$ because it had a more preferred option at that time, or it initially accepted UE $m$ but was later replaced by another UE in subsequent iterations. Thus, $m \notin \mathcal{W}_w(\mu(w) \cup {m})$ cannot be a false statement, implying that the matching $\mu$ is stable.
\rule{1.8mm}{1.8mm}

\begin{table*}[htbp]
\centering
\caption{Complexity Comparisons for the CSI Prediction of 100 RBs among Different Networks}
\label{tab:complexity_comparison}
\begin{tabular}{c|cc|cc}
\toprule\hline
\multirow{2}{*}{Model} & \multicolumn{2}{c|}{Space Complexity} & \multicolumn{2}{c}{Time Complexity} \\
\cline{2-5}
 & Number of Parameters & Memory Usage (MB) & Number of FLOPs & Computational Time (s) \\
\hline
Proposed CaFTRA & 43,975,233 &  167.8 & 3,192,893,440 & $5.32\times 10^{-5}$ \\
Independent Transformer Network & 275,731,300 & 1051.8 & 341,388,800 & $5.69\times 10^{-6}$ \\
Independent FCNN & 41,324,900 & 157.6 & 41,139,200 & $6.86\times 10^{-7}$ \\
FCNN & 38,510,948 & 146.9 & 38,490,112 & $6.42\times 10^{-7}$ \\
\toprule\hline
\end{tabular}
\end{table*}

\textbf{\emph{Theorem 2:}}
The M3-MAMA algorithm is guaranteed to terminate within limited iterations.

\textbf{\emph{Proof:}}
At each step, the algorithm only accepts allocations or exchanges that strictly improve the total system throughput. Since the number of possible allocations is finite and no allocation is repeated, the process must eventually reach a configuration where no further improvement is possible. This ensures that the algorithm will always terminate in a finite number of steps, regardless of the problem's convexity. Thus, M3-MAMA is guaranteed to converge.
\rule{1.8mm}{1.8mm}

\emph{Computational Complexity}:
The M3-MAMA algorithm employs two nested loops over $W$ BS-RB pairs. As the throughput evaluation has linear complexity, the overall computational complexity is $\mathcal{O}(W^2)$, which is polynomial and thus feasible for practical use. In contrast, exhaustive search has exponential complexity $\mathcal{O}(2^{MW})$, which is computationally prohibitive for large-scale networks.
The intended deployment scope of M3-MAMA is a localized multi-BS cooperative region, where the number of candidate UEs is naturally bounded. For larger-scale deployments, the service area can be partitioned into multiple geographical coordination regions,
each performing independent MAC-layer scheduling in parallel. This regional decomposition prevents
uncontrolled complexity growth and is consistent with practical clustered RAN operation.

\section{NUMERICAL ANALYSIS}

In this section, we detail the simulation setup and conduct extensive simulations to evaluate the effectiveness and fairness of the proposed CaFTRA in FD-RAN. Referring to 3GPP standard, the proposed feedback-free MIMO transmission, multi-BS association and multi-RB allocation are implemented on the Vienna 5G system-level simulator \cite{Vienna5GSLS}.

\subsection{Simulation Parameter Setting}

\begin{figure}[htbp]
\centering
		\includegraphics [width=3.5in]{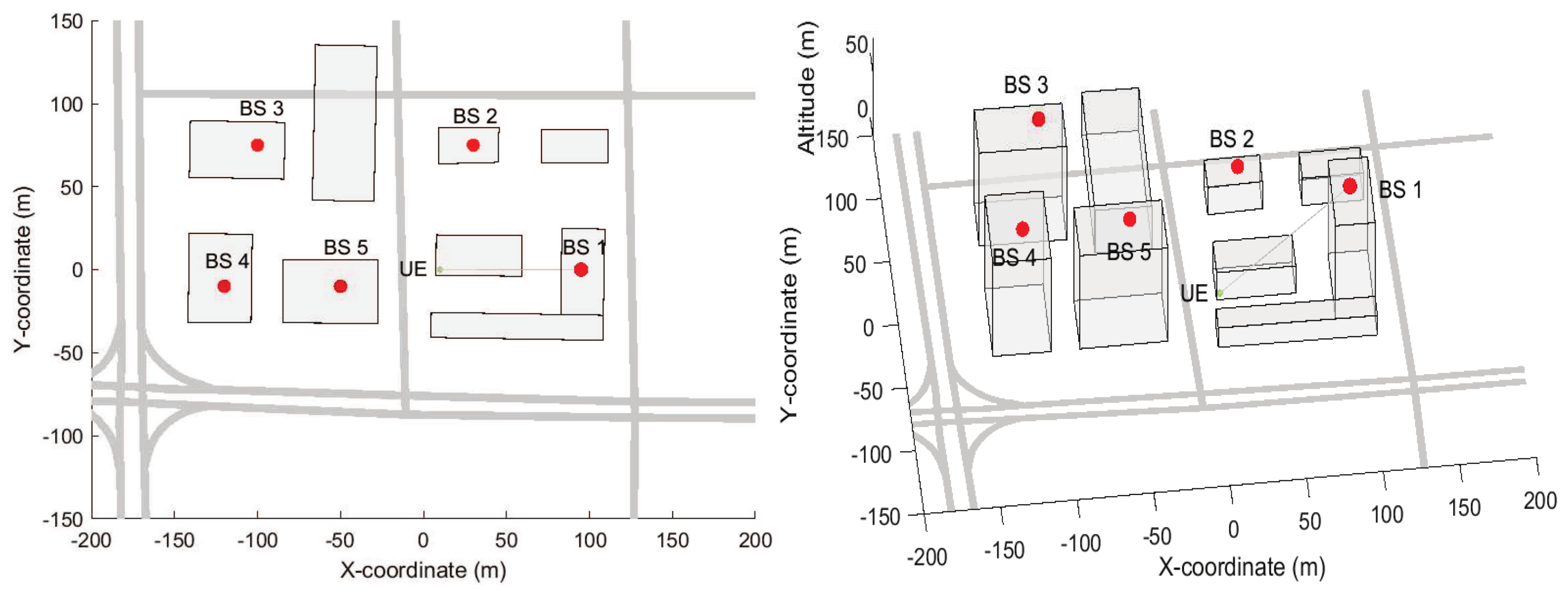}
\caption{2D (left) and 3D (right) Views of the Peng Cheng Laboratory Scenario in Vienna 5G System-Level Simulator.}\label{fig:PCL}
\end{figure}

Fig. \ref{fig:PCL} illustrates the simulation environment, i.e., Peng Cheng Laboratory, in the Vienna 5G system-level simulator.
It is a 400 m $\times$ 300 m urban area located within the geographical coordinates [113.93371-113.93792, 22.57494-22.57709] in OpenStreetMap, which is a typical Urban micro (UMi) scenario.
It includes 9 buildings and 5 micro-cell BSs placed on the rooftops.
Each BS is allocated 20 MHz bandwidth, i.e., 100 RBs.
Both 2D and 3D perspectives are provided in Fig. \ref{fig:PCL} to comprehensively depict the urban features of the scenario.

\subsection{Numerical Results}

In the proposed CSI prediction of CaFTRA framework, each input of an UE's geolocation is embedded into a 1024-dimension vector to predict the CSI of 100 RBs for a certain BS. The model consists of a 1-layer Transformer encoder and a 2-layer Transformer decoder.
To validate the effectiveness of modeling frequency-domain correlation, we construct the Independent Transformer Network as an ablation baseline by removing the correlation modeling components from CaFTRA.
We use an independent Transformer network to predict the CSI for each RB, and the embedding dimension is reduced to 256.
The independent model also comprises a 1-layer Transformer encoder and a 2-layer Transformer decoder, with similar structure to the proposed prediction model in CaFTRA but with fewer parameters.
Moreover, we add two additional CSI prediction baselines, namely Fully Connected Neural Networks (FCNN) and Independent FCNN, to provide a broader comparison beyond the independent Transformer baseline.
Specifically, the FCNN baseline uses one large fully connected neural network to jointly predict the CSI parameters over all 100 RBs, with hidden-layer dimensions \{2048, 2048, 2048, 6500\}.
In contrast, the Independent FCNN baseline consists of 100 independent small fully connected neural networks, each predicting the CSI of one RB, with hidden-layer dimensions \{256, 256, 256, 65\} for each subnetwork.
To ensure a fair comparison, we also present the corresponding model complexities, including the number of parameters, memory usage, FLOPs, and computational time.
As shown in Table \ref{tab:complexity_comparison}, the compared models are kept within a comparable order of parameter scale.
The reported computational latency corresponds to one BS-UE pair over 100 RBs.

The number of parameters and FLOPs can be obtained from the codes.
According to the parameter amount of the model, like \cite{{TF-CSI}}, the size of the memory can be calculated as:
\begin{equation}
\mbox{Memory}=\frac{4\times N_{parameter}}{1024^2},
\end{equation}
where $N_{parameter}$ represents the parameter amount of the network.
As the NVIDIA H100 GPU delivers up to 60 TFLOPS (tera FLOPS) of single-precision performance, resulting in estimation computation time (i.e., the inference time of the CSI for each BS).
As shown in Table \ref{tab:complexity_comparison}, the proposed CaFTRA model significantly reduces the space complexity compared to the Independent Transformer Network. Specifically, CaFTRA requires only 44 million parameters and 168 MB of memory, which are approximately 84\% lower, than those of the baseline model.
In terms of time complexity, although CaFTRA involves more FLOPs due to its correlation-aware architecture (3193 M vs. 341 M), its overall computational time remains within the same order of magnitude, indicating a highly parallelizable and hardware-friendly structure.

The proposed CaFTRA framework is to be deployed in edge cloud or AI-RAN infrastructure \cite{AIRAN,AIRAN2} rather than per-BS.
Since CSI prediction in CaFTRA is geolocation-driven, inference is only required when the UE geolocation changes beyond the update granularity, rather than at each transmission time interval (TTI).
In practical deployment, multiple BS-UE pairs can be processed in parallel through batched inference at the edge cloud.
Therefore, NVIDIA H100 GPU is used here only as a reference for the inference time calculation.


1) \emph{Data Generation and CSI Heatmap}

\begin{figure}[h]
    \centering
    \subfigure[Heatmap of RI value.]{
        \includegraphics[width=0.43\linewidth]{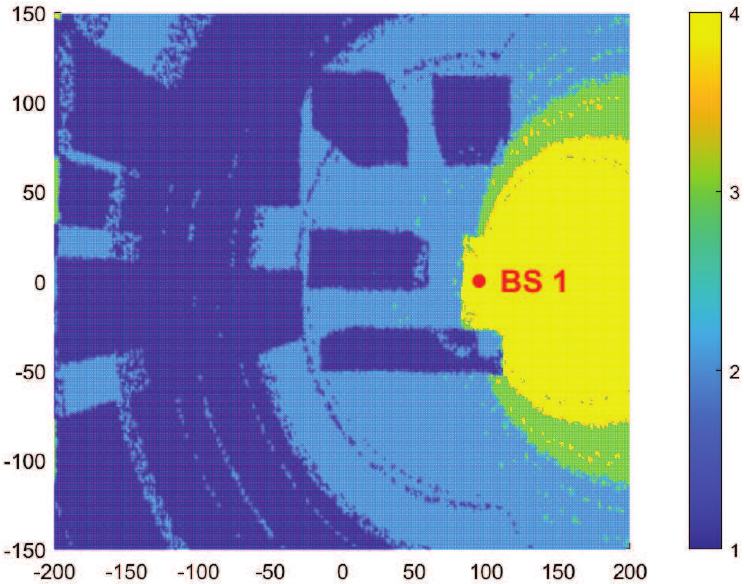}
    }
    \hspace{0.05\linewidth}
    \subfigure[Heatmap of CQI 1 value.]{
        \includegraphics[width=0.43\linewidth]{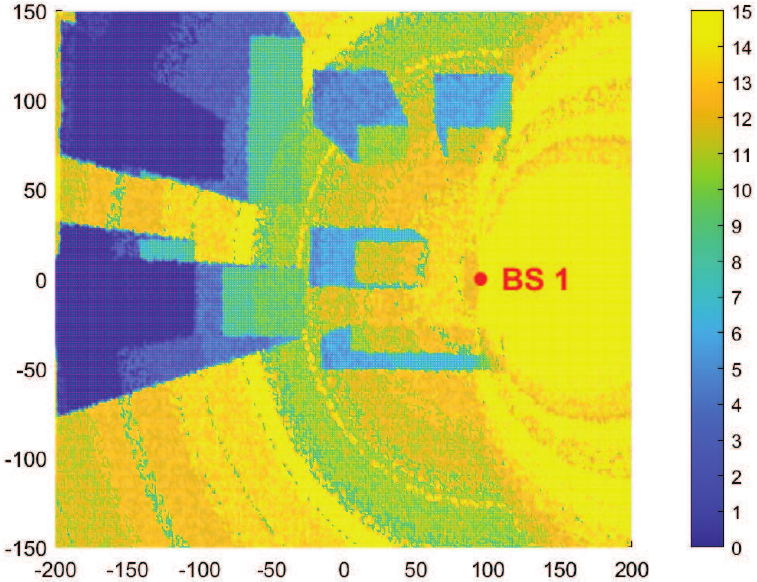}
    }

    \vspace{0.05\linewidth}
    \subfigure[Heatmap of CQI 2 value.]{
        \includegraphics[width=0.43\linewidth]{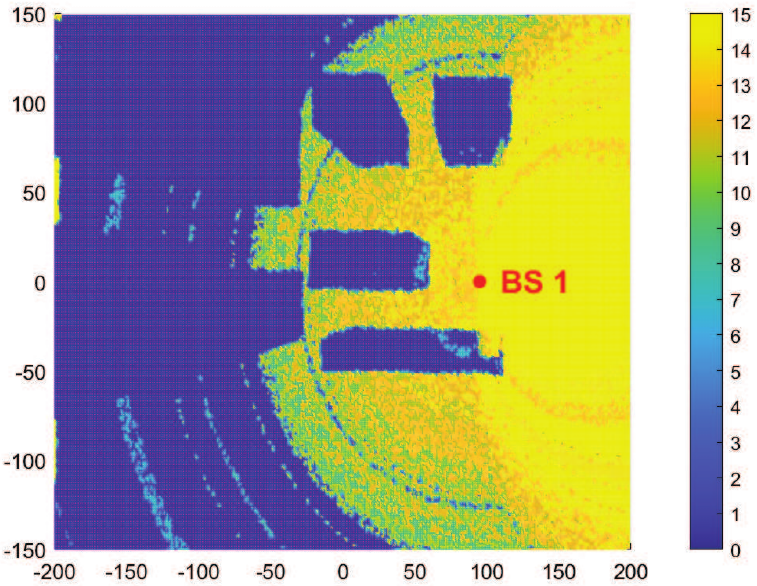}
    }
    \hspace{0.05\linewidth}
    \subfigure[Heatmap of PMI value.]{
        \includegraphics[width=0.43\linewidth]{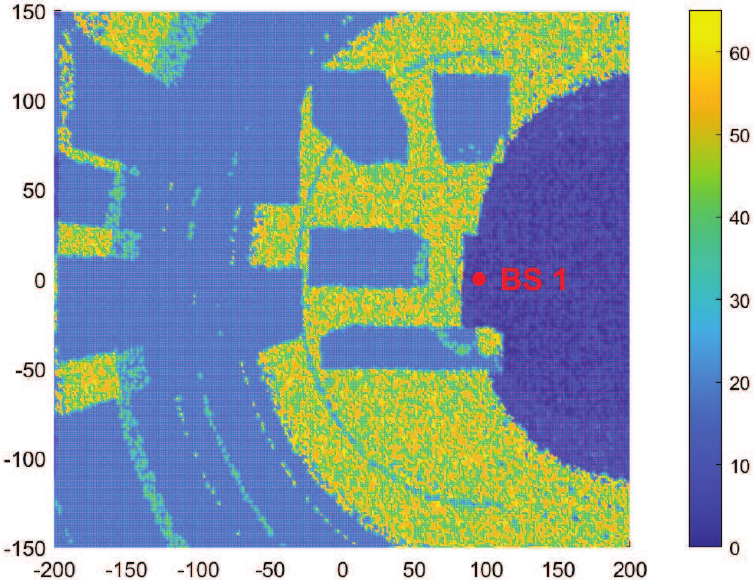}
    }

    \caption{Heatmap of historical CSI in training dataset.}
    \label{fig:heatCSI}
\end{figure}

\begin{table}[htbp]
\footnotesize
\renewcommand\arraystretch{1}
\begin{center}
\caption{MAIN SIMULATION PARAMETERS}\label{Simulation}
\begin{tabular}{c|c}
\toprule\hline
Parameters  & Value   \\
\hline
MIMO Type & SU-MIMO \\
Frame Structure & FDD\\
Waveform & OFDM \\
Codebook  & 3GPP Type I\\
Carrier Frequency & 3.5 GHz\\
Total Bandwidth  & 100 MHz\\
Number of RB & 500 \\
Feedback Delay & 3 ms\\
Micro-Cell BS Antenna Panel & Single Polarization (6, 1)\\
Number of UE Antennas & 4 \\
Channel Model  & 3GPP TR 25.890\\
\toprule\hline
\end{tabular}
  \end{center}
\end{table}

The main simulation parameters are listed in Table \ref{Simulation}.
For the proposed LQTN-based CSI map, a historical CSI dataset with 50,000 randomly sampled user geolocations is first generated in Vienna 5G system-level simulator, as the labeled training dataset.
Another 5,000 randomly generated user geolocations are used as test data to evaluate the CSI prediction accuracy and MIMO transmission performance in Vienna 5G system-level simulator.
In practical deployment, such historical CSI samples can be collected from current BS services at different geolocations. Moreover, users can periodically feedback CSI to the control BS, and these measurements are aggregated by the edge cloud as training data for the CSI prediction.
By periodically refreshing and retraining the CaFTRA model with newly collected CSI samples, the system can effectively mitigate the risk of CSI staleness and maintain robustness against environment changes or CSI aging.

Fig. \ref{fig:heatCSI} (a)-(d) illustrates the heatmaps of historical CSI parameters, namely RI, CQI 1, CQI 2, and PMI, coming from the training dataset consisting of 50,000 geolocations. Since CSI is essentially a quantized representation of MIMO transmission channel quality ranging from poor to good, these heatmaps provide an intuitive view of the spatial distribution of different CSI.
It can be observed that CSI varies gradually with respect to geolocations, leveraging user geolocation as the basis for CSI prediction is feasible and necessary.
Moreover, each CSI tends to form distinct regional patterns across the coverage area rather than fluctuating sharply on a fine-grained scale.
This indicates that the CSI parameters are relatively insensitive to small prediction error of user precise geolocation, and instead exhibit more regionally consistent behaviors.

2) \emph{CSI Prediction Performance Against Geolocation Error and Different Models}

\begin{figure}[H]
\centering
\includegraphics [width=3.2in]{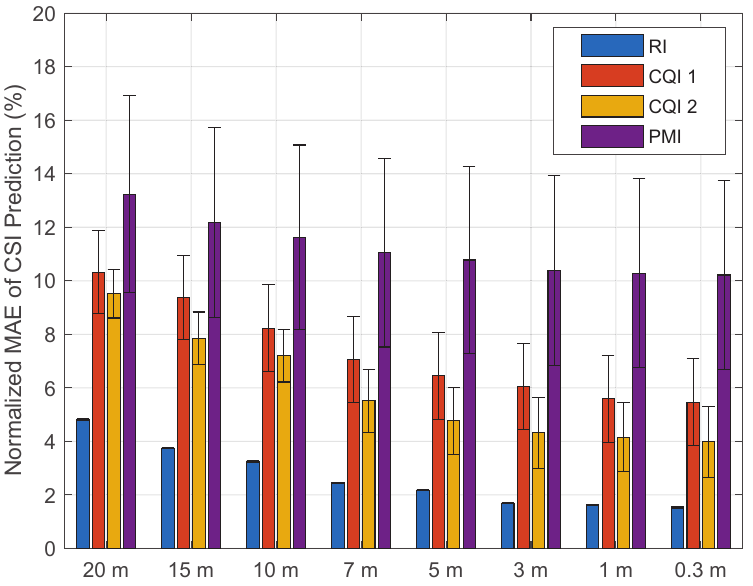}
\centering
\caption{Normalized MAE of CSI Prediction Using CaFTRA Among Different Geolocation Error $r_{95}$ with 95\% Confidence Level in BS 1.}
\label{R1-error_loc}
\end{figure}

To ensure the practicality of the UE geolocation uncertainty, we referred to existing 3GPP specifications and studies on UE positioning accuracy to guide our simulation settings.
In 3GPP TR 38.855 \cite{3GPPTR38.855}, Table 8.1.2.11-2 shows that the horizontal positioning error of Urban micro (UMi) scenario is 2.6 meter with 95\% percentile using downlink time difference of arrival (DL-TDOA) positioning.
In 3GPP TS 22.261 \cite{3GPPTS22.261}, Table 7.3.2.2-1 defines the positioning service level 1-7 for the cellular network, where the accuracy $r_{95}$ ranges from 10 meter to 0.2 meter  with 95\% confidence level.
3GPP TS 38.305 \cite{3GPPTS38.305} further enhances the UE positioning framework by supporting multiple standard UE positioning methods.
Referring to \cite{Gauss}, we model the geolocation error on each coordinate as a one-dimensional Gaussian random variable.

We vary $r_{95}$ from 20 m to 0.3 m to cover a wide range of practical positioning accuracies.
Fig. \ref{R1-error_loc} shows the normalized mean absolute error (MAE) of CSI prediction in BS 1 using CaFTRA.
As the geolocation accuracy improves, the prediction error gradually decreases.
More importantly, even under relatively large positioning errors (e.g., $r_{95}=20$ m), the prediction error remains within a moderate range. For instance, the normalized MAE of RI prediction is below 5\%, and the errors for CQI and PMI remain bounded.
It proves that the proposed CaFTRA framework does not rely on extremely precise UE positioning and remains applicable under practical network positioning capabilities.
We conduct the following simulations under ideal UE geolocation input.

\begin{figure}[H]
\centering
\includegraphics [width=3.2in]{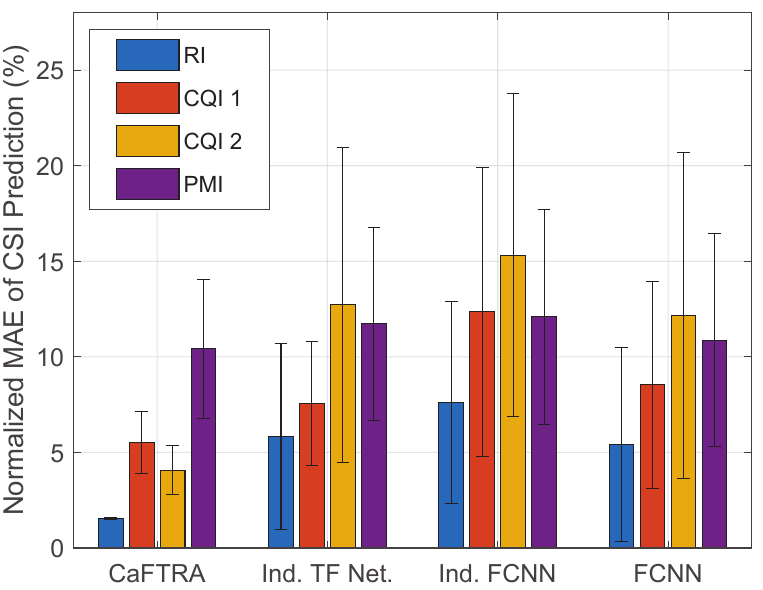}
\centering
\caption{Normalized Mean Absolute Error of CSI Prediction Comparison Among CaFTRA, Independent Transformer Network (Ind. TF Net.), Independent FCNN (Ind. FCNN), and FCNN in BS 1.}
\label{R2-FCNN}
\end{figure}

Fig. \ref{R2-FCNN} presents the normalized Mean Absolute Error (MAE) of CSI prediction at BS 1 for frequency-domain Correlation-aware Feedback-free MIMO Transmission and Resource Allocation (CaFTRA) and three CSI prediction baselines.
The results show that the proposed CaFTRA achieves the best prediction performance among all baselines. Moreover, the FCNN baseline outperforms the Independent FCNN baseline, which further supports our claim that exploiting frequency-domain correlation across RBs is beneficial for improving CSI prediction accuracy.

3) \emph{CSI Prediction Against Different Scenario and Carrier Band}

\begin{figure}[H]
\centering
\includegraphics [width=3.5in]{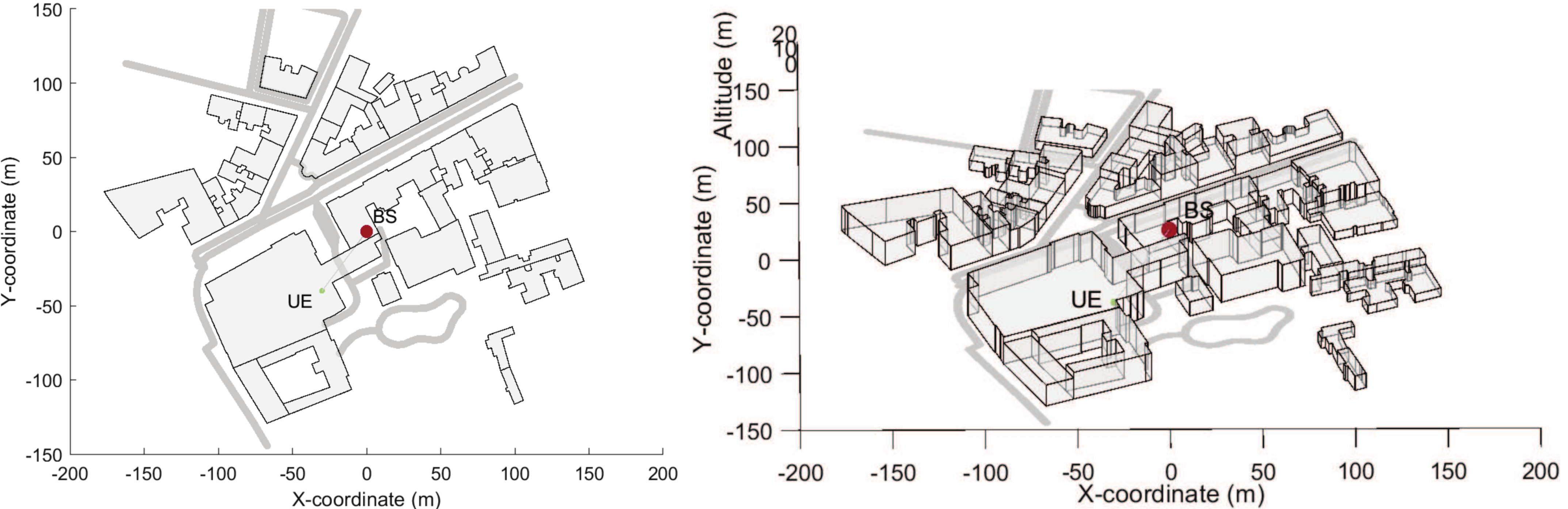}
\centering
\caption{2D (left) and 3D (right) Views of the Shopping Mall Scenario in Vienna 5G System-Level Simulator.}
\label{map_mall}
\end{figure}

To further examine the robustness of the proposed CaFTRA framework beyond the original Peng Cheng Laboratory scenario, we additionally consider a classical Shopping Mall scenario, in the Vienna 5G system-level simulator based on the OpenStreetMap with geographical coordinates
[48.1955904-48.1973271, 16.3690209-16.3718059] at Vienna, Austria.
As shown in Fig. \ref{map_mall}, the BS is placed on the rooftop of the central building.
We evaluate the CSI prediction performance under two frequency bands, 3.5 GHz and 7 GHz, to examine both environmental and frequency-domain variation.
The purpose of this experiment is not to claim universal zero-shot generalization across arbitrary maps, but to verify that CaFTRA remains effective as a deployment-specific and retrainable framework across different environments and carrier bands.

\begin{figure}[h]
    \centering
    \subfigure[Heatmap of RI value.]{
        \includegraphics[width=0.43\linewidth]{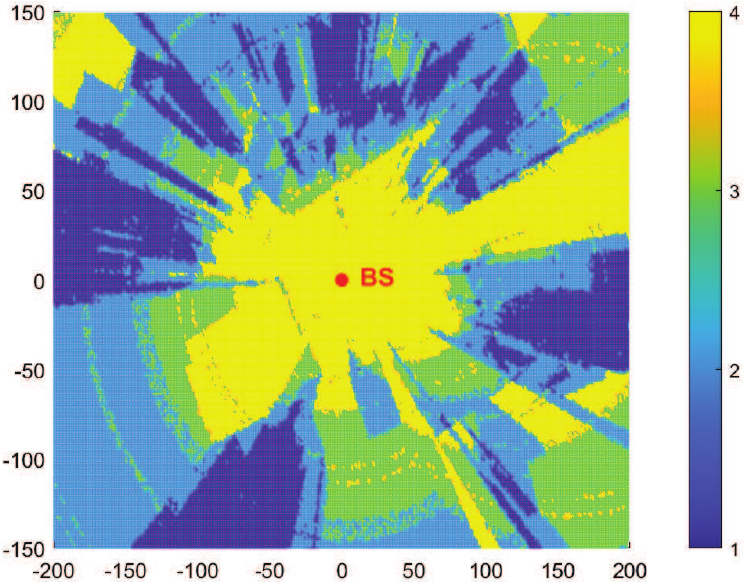}
    }
    \hspace{0.05\linewidth}
    \subfigure[Normalized Mean Absolute Error of CSI Prediction.]{
        \includegraphics[width=0.43\linewidth]{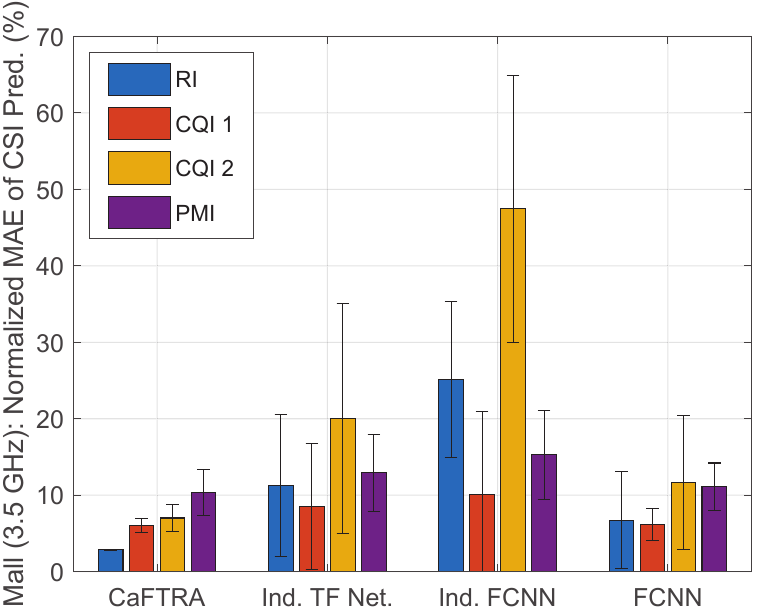}
    }
    \caption{Heatmap of historical RI in training dataset and CSI prediction results at Shopping Mall Scenario with BS operating at 3.5 GHz band.}
    \label{fig:mall35G}
\end{figure}

Compared with the original Peng Cheng Laboratory scenario in Fig. \ref{fig:heatCSI}, the RI heatmap distribution in the Shopping Mall scenario, i.e., Fig. \ref{fig:mall35G}(a) is clearly different, confirming that CSI characteristics are strongly scenario-dependent.
We therefore retrain all compared models in this new environment using 50,000 randomly generated UE geolocations and evaluate them on 5,000 new geolocations.

Fig. \ref{fig:mall35G}(b) shows that CaFTRA still achieves the best prediction performance among all compared models.
Although the absolute accuracy is slightly lower than in the original scenario, this is consistent with the higher propagation complexity of the Shopping Mall environment.
In addition, the Transformer-based models show a larger advantage over FCNN-based models in this more complex setting, indicating that the proposed architecture captures more informative CSI features under complicated propagation conditions.

\begin{figure}[h]
    \centering
    \subfigure[Heatmap of RI value.]{
        \includegraphics[width=0.43\linewidth]{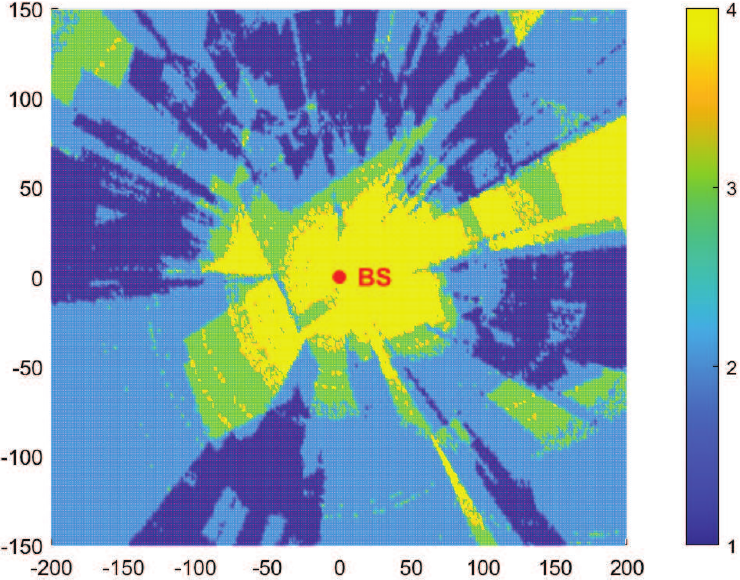}
    }
    \hspace{0.05\linewidth}
    \subfigure[Normalized Mean Absolute Error of CSI Prediction.]{
        \includegraphics[width=0.43\linewidth]{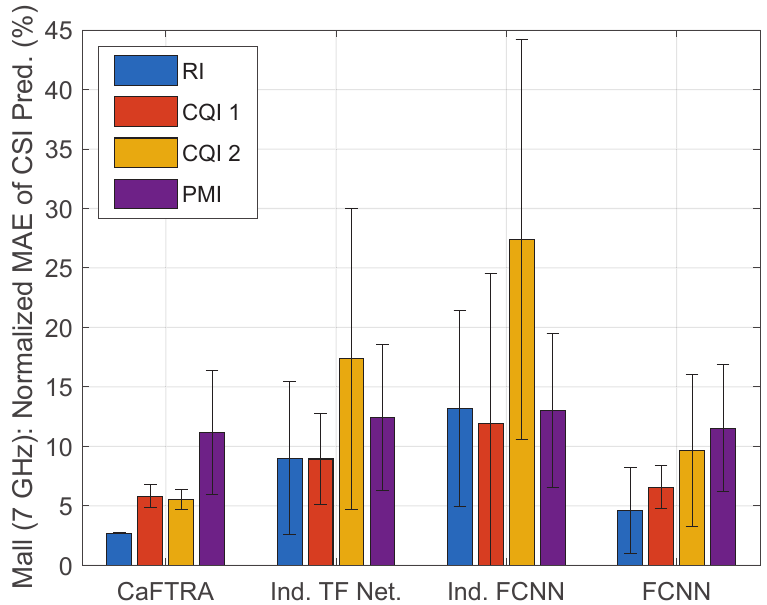}
    }
    \caption{Heatmap of historical RI in training dataset and CSI prediction results at Shopping Mall Scenario with BS operating at 7 GHz band.}
    \label{fig:mall7G}
\end{figure}

We further evaluate the Shopping Mall scenario at 7 GHz. The RI heatmap, i.e., Fig. \ref{fig:mall7G}(a), shows further quality degradation compared with the 3.5 GHz case, which is consistent with the stronger path loss at the higher carrier frequency.

Nevertheless, after retraining on the new band-specific data, CaFTRA again achieves the best CSI prediction performance among all compared methods. Moreover, models that jointly predict CSI across all 100 RBs consistently outperform models that predict each RB independently, further validating the importance of explicitly exploiting frequency-domain correlation.


4) \emph{Comparison with 5G Feedback-Based MIMO}

\begin{figure}[h]
    \centering
    \subfigure[5G CLSM.]{
        \includegraphics[width=0.43\linewidth]{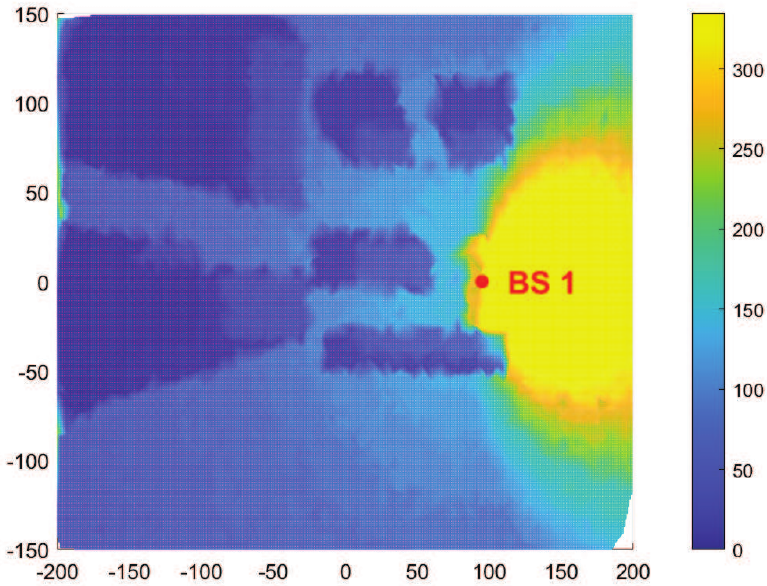}
    }
    \hspace{0.05\linewidth}
    \subfigure[CaFTRA-based FD-RAN.]{
        \includegraphics[width=0.43\linewidth]{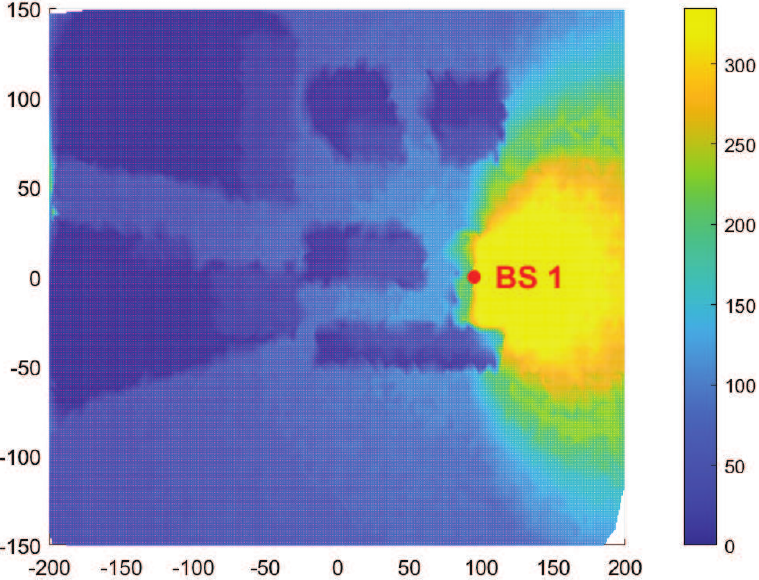}
    }
    \caption{Throughput (Mbps) Heatmap Comparision of 5G CLSM and CaFTRA in Test Data under BS 1.}
    \label{fig:heat}
\end{figure}

Fig. \ref{fig:heat} illustrates the throughput heatmaps of the test data (i.e., 5,000 users) for BS 1 under two different MIMO transmission methods, specifically for static scenario.
Similar observations hold for the other four BSs, and therefore BS 1 is taken as an example here.
Fig. \ref{fig:heat} (a) represents the throughput distribution using 5G CLSM, i.e., a feedback-based 5G MIMO transmission method, while Fig. \ref{fig:heat} (b) shows the throughput distribution using the proposed CaFTRA without feedback.
The shaded regions in the figure indicate the presence of building-induced obstructions.
It demonstrates that the proposed CaFTRA achieves comparable throughput compared to 5G CLSM, while eliminating the CSI feedback.
From a macroscopic perspective, the throughput distributions of 5G CLSM and CaFTRA-based FD-RAN are generally consistent. This implies that the CSI prediction mechanism of CaFTRA can effectively infer the users' CSI characteristics in the spatial domain, since the geolocations in the test data were not included in the training dataset.

\begin{figure}[H]
\centering
\includegraphics [width=3.2in]{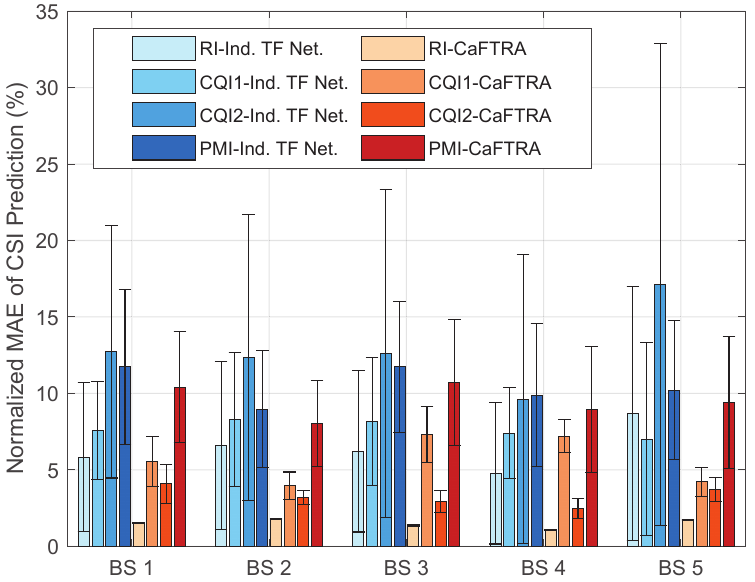}
\centering
\caption{Normalized Mean Absolute Error of CSI Prediction Comparison between Independent Transformer Network (Independent TF) and  CaFTRA Across BSs 1 to 5.}
\label{fig:CSI_pre}
\end{figure}

Fig. \ref{fig:CSI_pre} illustrates the normalized MAE of CSI prediction across BS 1 to BS 5 for the four CSI components, namely RI, CQI 1, CQI 2, and PMI. Two approaches are compared: the proposed CaFTRA framework and a baseline method using independent transformer networks, where each RB is predicted separately.
The results show that CaFTRA consistently achieves lower MAE values across all CSI components. The superior performance of CaFTRA is attributed to its ability to exploit frequency-domain correlations among adjacent RBs through joint learning, rather than treating each CSI of RB in isolation.
In addition, CaFTRA exhibits smaller error variances, as reflected by the shorter error bars, indicating enhanced prediction stability and robustness. These results validate the effectiveness of our frequency correlation-aware, feedback-free design in improving CSI prediction accuracy.

\begin{figure}[H]
\centering
\includegraphics [width=3.2in]{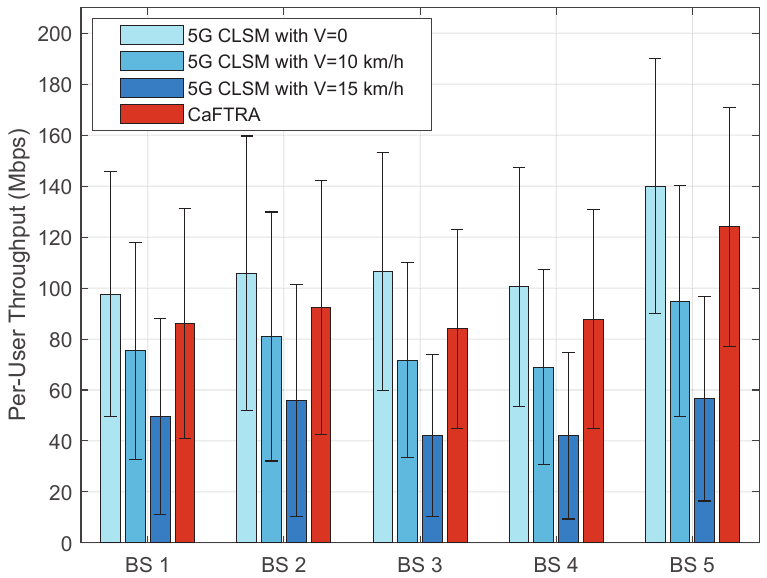}
\centering
\caption{Per-User Throughput Comparison for 5G CLSM and CaFTRA-based FD-RAN Across BS 1 to BS 5.}\label{fig:loss}
\end{figure}

Fig. \ref{fig:loss}  illustrates the throughput across BS 1 to 5 for 5G Feedback-Based MIMO (i.e., CLSM) and the proposed CaFTRA framework under varying user mobility conditions for the 5,000 users in test data.
With the setting of 3 ms feedback delay, CaFTRA's performance is minimally affected under low-speed scenarios, as predicting geoglocations is significantly more accurate and reliable than predicting CSI changes.
This delay is incurred by the full chain of operations: the BS first transmits pilot symbols, the UE estimates the instantaneous CSI, and then feeds it back to the BS. Thus, the chosen 3 ms reflects an optimistic assumption for CLSM, providing its theoretical best-case performance under feedback.
In contrast, the proposed CaFTRA does not rely on instantaneous CSI feedback at all, and it only requires user geolocations, which can be predicted more reliably.

For 5G CLSM, the throughput outperforms the proposed feedback-free method by less than 14\% when users are static (speed = 0). However, At a user velocity of 10 km/h, CaFTRA outperforms 5G CLSM by around 20\% in terms of per-user throughput. At 15 km/h, CaFTRA surpasses 5G CLSM by 93\%, showcasing its superiority in mobility scenarios.
These results demonstrate that in handling user mobility, feedback-free MIMO transmission can achieve consistent and reliable performance by leveraging accurately predicted geolocation, even under significant transmission delays. This makes FD-RAN a promising solution for real-time applications in high-mobility environments.

As a conclusion of this part, for the feedback-based MIMO of 5G, it heavily relies on accurate and real-time CSI feedback to maintain optimal transmission performance.
This dependency results in significant spectral overhead and causes severe performance degradation in high-mobility scenarios due to the rapidly changing channel conditions.
The proposed feedback-free MIMO transmission solution (i.e., CaFTRA) eliminates the need for real-time CSI feedback by leveraging user geolocation that are easier to predict and obtain, significantly reducing spectrum resource consumption while ensuring better reliability and superior performance, particularly in high-mobility environments.

5) \emph{Performance Improvements by the Extended Coverage of FD-RAN}

We will further verify that, beyond dynamic scenarios, FD-RAN also demonstrates significant advantages over 5G in static environments.
This is due to that the physical decoupling of BSs significantly expands the downlink coverage area.
In FD-RAN, more flexible multi-BS collaborative transmission and resource scheduling can be achieved, further enhancing its overall performance and adaptability.

\begin{figure}[H]
\centering
\includegraphics [width=3.2in]{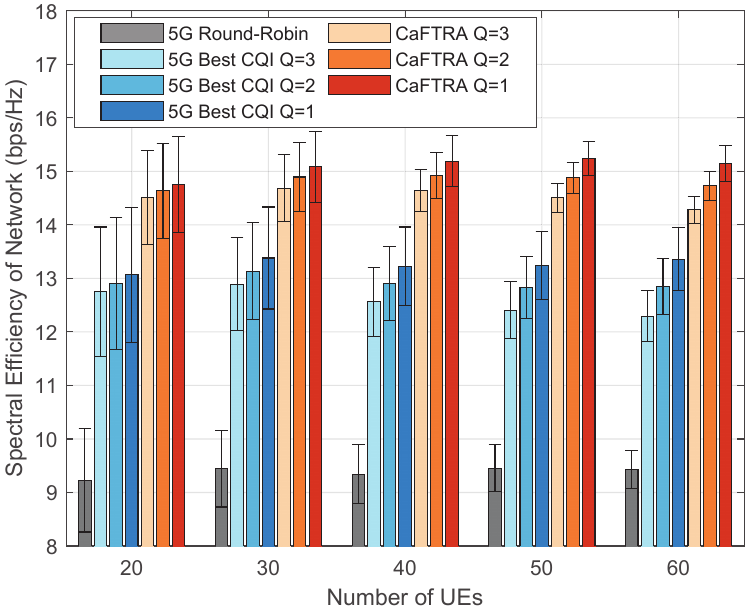}
\centering
\caption{Spectral Efficiency Comparison of 5G Round-Robin, 5G Best CQI, and CaFTRA-based FD-RAN under Different $Q$ Levels.}\label{fig:Thr}
\end{figure}

In the Vienna 5G system-level simulator \cite{Vienna5GSLS}, two scheduling methods are implemented: \textbf{5G Round-Robin} scheduler and \textbf{5G Best CQI} scheduler.
The round robin scheduler schedules the users one after the other in a row. When all users have been assigned resources, the first user is scheduled again and so on.
In case of the best CQI scheduler, the user with the highest CQI value calculated bay the feedback is allocated to each RB.
In the following simulations, we evaluated the MAC-layer resource allocation results for UE number ranging from 20 to 60 under constraints of $Q = 1, 2, 3$, generating 100 random UE geolocations to compute the mean values and plot error bars (i.e., sample variance).
Since the proposed MAC scheduler, i.e., M3-MAMA, is formulated under an SE-oriented utility, Best-CQI is used as the main objective-aligned reference, while Round-Robin is included as an additional representative scheduler for comparison.

Fig. \ref{fig:Thr} shows the comparison of spectral efficiency for the proposed CaFTRA framework against two commonly used 5G scheduling algorithms: Round-Robin and Best CQI. These algorithms prioritize fairness and spectral efficiency, respectively. The evaluation is conducted for static users (speed = 0) with the number of users ranging from 20 to 60 under varying $Q$ guarantees ($Q = 1, 2, 3$).
It can be observed that CaFTRA consistently outperforms both 5G Round-Robin and 5G Best CQI across all QoS levels and user numbers. Specifically, CaFTRA achieves a 60\% improvement in spectral efficiency compared to 5G Round-Robin and about 15\% improvement compared to 5G Best CQI. The performance gain is particularly notable under higher $Q$ constraints ($Q = 3$), where CaFTRA demonstrates its ability to better allocate resources while ensuring stringent service quality requirements.
The reported spectral efficiency is a system-level result under the considered UMi-like micro-cell SU-MIMO cooperative setup. Its absolute value should therefore be interpreted within this specific simulation context, while the main message of this figure is the relative gain of CaFTRA over the representative 5G baselines under the same deployment assumptions.

\begin{figure}[H]
\centering
\includegraphics [width=3.2in]{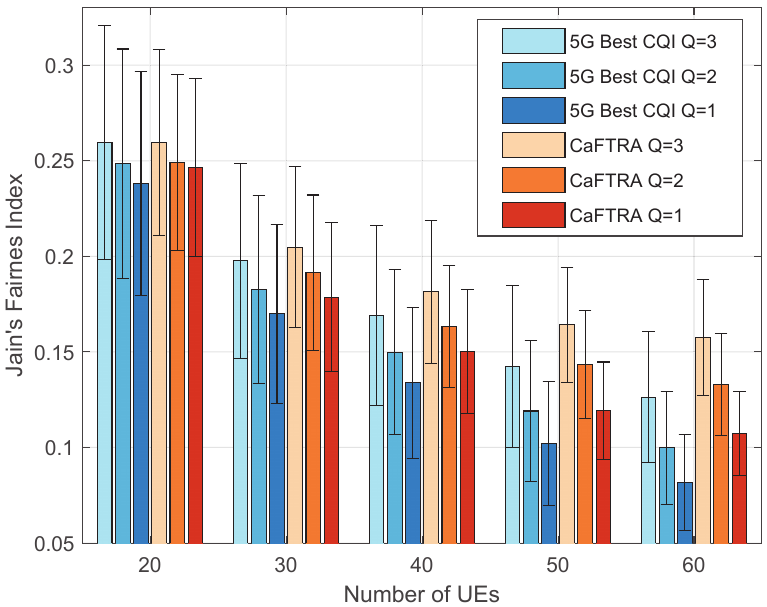}
\centering
\caption{Jain's Fairness Index Comparison of 5G Best CQI and CaFTRA-based FD-RAN.}\label{fig:JAIN}
\end{figure}

As the number of users increases, the impact of $Q$ values on spectral efficiency becomes more significant. This is because the total number of RBs remains fixed, and with more users, the RBs available for flexible scheduling decrease, leading to a reduction in performance gains as $Q$ increases. The superior performance of CaFTRA stems from its learning-oriented feedback-free transmission approach, which allows for more efficient multi-BS association and multi-RB allocation. This adaptability enables CaFTRA to effectively utilize spectral resources, even as the number of users increases, while maintaining higher spectral efficiency than conventional 5G methods. These results highlight the potential of CaFTRA as an efficient resource management solution in 6G and beyond.

Fig. \ref{fig:JAIN} illustrates the \textbf{Jain's Fairness Index} for 5G Best CQI and the proposed CaFTRA under varying numbers of users and different $Q$ guarantees ($Q = 1, 2, 3$). Jain's Fairness Index \cite{jain} is a widely recognized metric for evaluating the fairness of resource scheduling algorithms in multi-user networks, where higher values indicate better fairness.
For any given set of user throughputs $(x_1, x_2, \cdots, x_n)$, the Jain's Fairness Index is calculated as follows:
\begin{equation}
f(x_1, x_2, \cdots, x_n)=\frac{(\sum_{i=1}^n x_i)^2}{n\sum_{i=1}^n x_i^2}
\end{equation}

From Fig. \ref{fig:JAIN}, it is evident that CaFTRA consistently achieves significantly higher fairness compared to 5G Best CQI across all user densities and $Q$ levels. The fairness gap becomes more pronounced as the number of users increases, highlighting the ability of CaFTRA to maintain equitable resource allocation even under high user density. Moreover, for both algorithms, fairness improves as the $Q$ level increases, since larger $Q$ values provide more flexibility in balancing resource allocation among users.

\begin{figure}[H]
\centering
\includegraphics [width=3.2in]{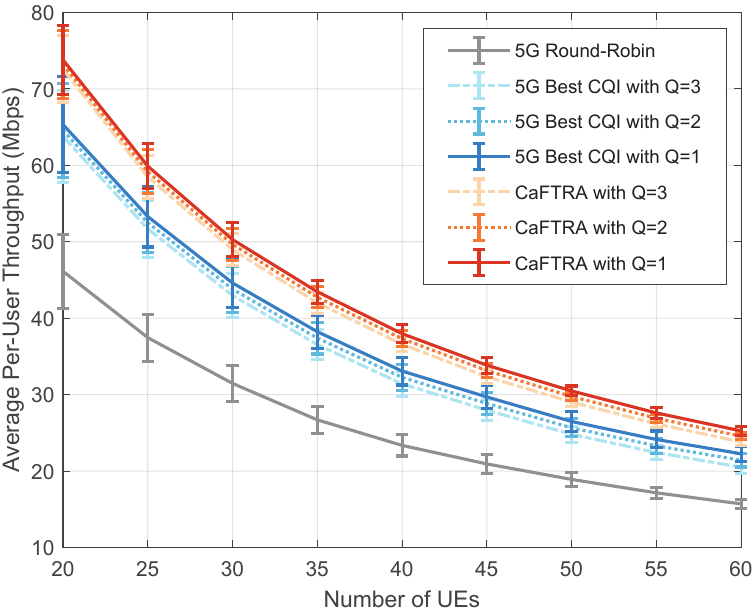}
\centering
\caption{Average Per-User Throughput Comparison of 5G Round-Robin, 5G Best CQI, and CaFTRA-based FD-RAN.}\label{fig:AVG}
\end{figure}

Fig. \ref{fig:AVG} compares the average per-user throughput across three scheduling algorithms: 5G Round-Robin, 5G Best CQI, and the proposed CaFTRA under varying numbers of users, ranging from 20 to 60.
CaFTRA achieves higher per-user throughput under all tested QoS levels ($Q = 1, 2, 3$), maintaining consistent superiority as the number of users increases.
It is important to note that the simulation does not account for the communication resource savings achieved by eliminating CSI feedback in CaFTRA. Additionally, we set the feedback delay for both CaFTRA and 5G algorithms as the same (3 ms), further emphasizing the inherent efficiency of CaFTRA.

The superior performance of CaFTRA stems from its learning-oriented feedback-free transmission approach, which eliminates the dependence on real-time CSI feedback and instead leverages more predictable geolocation. This allows for more efficient multi-BS coordination and resource scheduling, resulting in significant throughput gains even as the number of users increases. These findings highlight the robustness and adaptability of CaFTRA, making it a compelling solution for scenarios with high user density and QoS requirements.

\begin{figure}[H]
\centering
\includegraphics [width=3.2in]{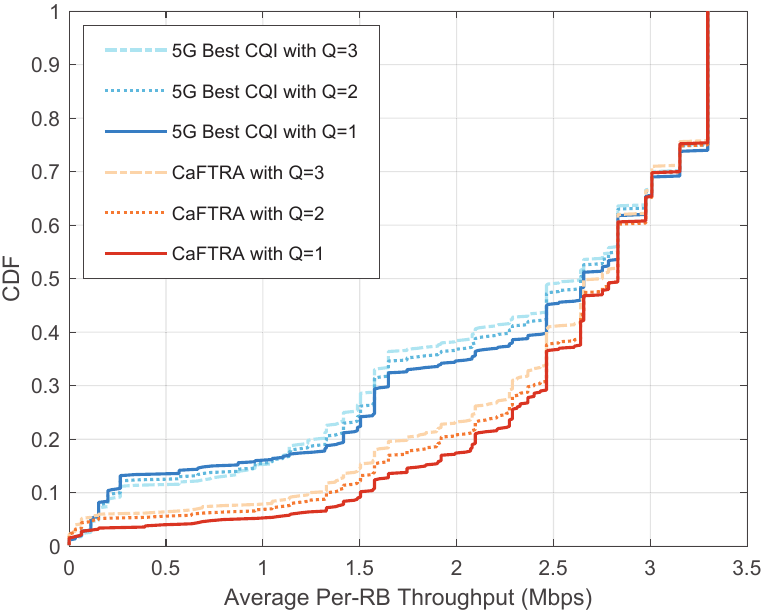}
\centering
\caption{CDF of Per-RB Throughput for 5G Best CQI and CaFTRA-based FD-RAN.}\label{fig:RB-CDF}
\end{figure}

Fig. \ref{fig:RB-CDF} presents the cumulative distribution function (CDF) of the per-RB throughput for 5G Best CQI and the proposed CaFTRA in FD-RAN under 30 UEs with different $Q$ levels ($Q = 1, 2, 3$).
The figure shows that CaFTRA consistently achieves higher per-RB throughput compared to 5G Best CQI across all QoS levels.
The superior performance of CaFTRA is attributed to its fundamental design advantages in FD-RAN. By decoupling UL and DL BSs, FD-RAN significantly expands the DL coverage area. This enables each DL BS to select the optimal UE for transmission on each RB, maximizing spectral efficiency. Additionally, CaFTRA supports collaborative transmission, allowing each UE to simultaneously benefit from RBs assigned by multi-BSs, further enhancing throughput performance.

The results highlight that the learning-oriented feedback-free design of CaFTRA not only eliminates the need for real-time CSI feedback but also leverages the structural flexibility of FD-RAN to achieve significant performance gains at the per-RB level, making it a promising solution for future wireless communication networks.

\begin{figure}[H]
\centering
\includegraphics [width=3.2in]{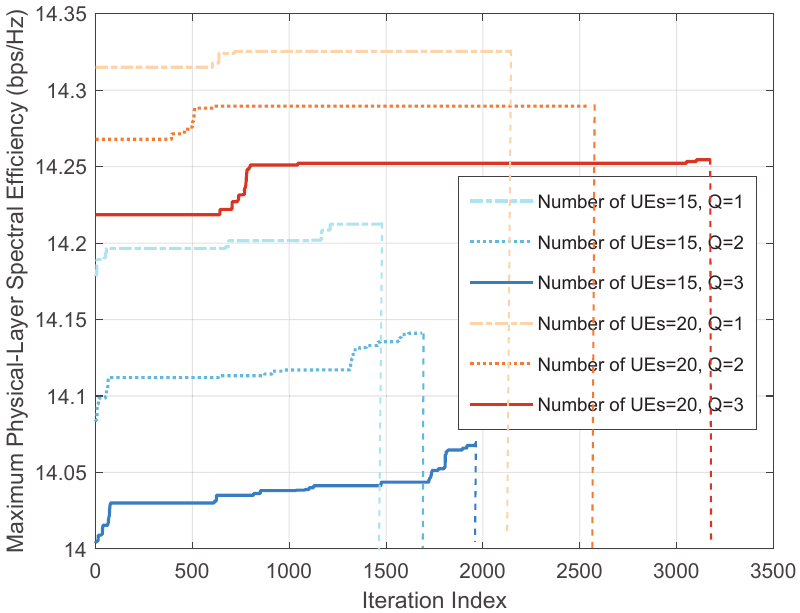}
\centering
\caption{Convergence of Exchange-Matching Iterations in M3-MAMA.}\label{fig:Converge}
\end{figure}

Fig. \ref{fig:Converge} illustrates the convergence of the Maximum Physical-Layer Spectral Efficiency in the CaFTRA as the number of exchange-matching iterations increases. The results are presented for two user densities (15 and 30 users) under different $Q$ levels ($Q = 1, 2, 3$).
From the figure, it is evident that as the number of users or the QoS level increases, the algorithm requires more iterations to converge. This is due to the increased number of optimization variables and constraints that need to be satisfied when accommodating more users or stricter QoS requirements. For instance, at $Q = 3$, the spectral efficiency improves more gradually, requiring a higher number of iterations compared to lower QoS levels.

As a conclusion of this part, for the MAC-layer resource allocation in 5G, it requires real-time CSI feedback and is limited by the single-connection mode and BS coverage, hindering optimal resource scheduling and multi-BS collaboration.
For the CaFTRA-based FD-RAN, the decoupled uplink and downlink significantly expand the DL BS coverage, enabling broader multi-BS and multi-RB cooperation.
This increases scheduling complexity but also brings performance gains.

\section{CONCLUSION}

In this paper, we have proposed the CaFTRA framework for FD-RAN, addressing critical limitations associated with channel information feedback-based MIMO transmission in cellular networks. We have introduced a Learnable Queries-driven Transformer Network, enabling frequency-domain correlation-aware CSI prediction across RBs, and the feedback-free MIMO transmission at the PHY layer based solely on UE geolocation.
Moreover, to fully explore the possibility of such extensive coverage with multi-BS cooperation, we have developed a low-complexity many-to-one matching algorithm for flexible multi-BS association and multi-RB allocation, and proved the convergence to stable matching within limited iterations.
Simulation results have shown that the proposed CaFTRA algorithm could achieve significant improvement in spectral efficiency compared to conventional 5G networks, revealing its effectiveness in improving performance and resource utilization in both static and high-mobility scenarios.
These findings have demonstrated that CaFTRA effectively addresses the critical challenges associated with CSI feedback overhead and resource allocation scalability.
The current study focuses on the SU-MIMO setting as a first step. Future work includes MU-MIMO-aware scheduling, and power control under feedback-free mode.


\ifCLASSOPTIONcompsoc
\else
\fi

\nocite{*}
\bibliographystyle{IEEEtran}
\bibliography{refernces_3C}

\begin{IEEEbiography}[{\includegraphics[width=1in,height=1.25in,clip,keepaspectratio]{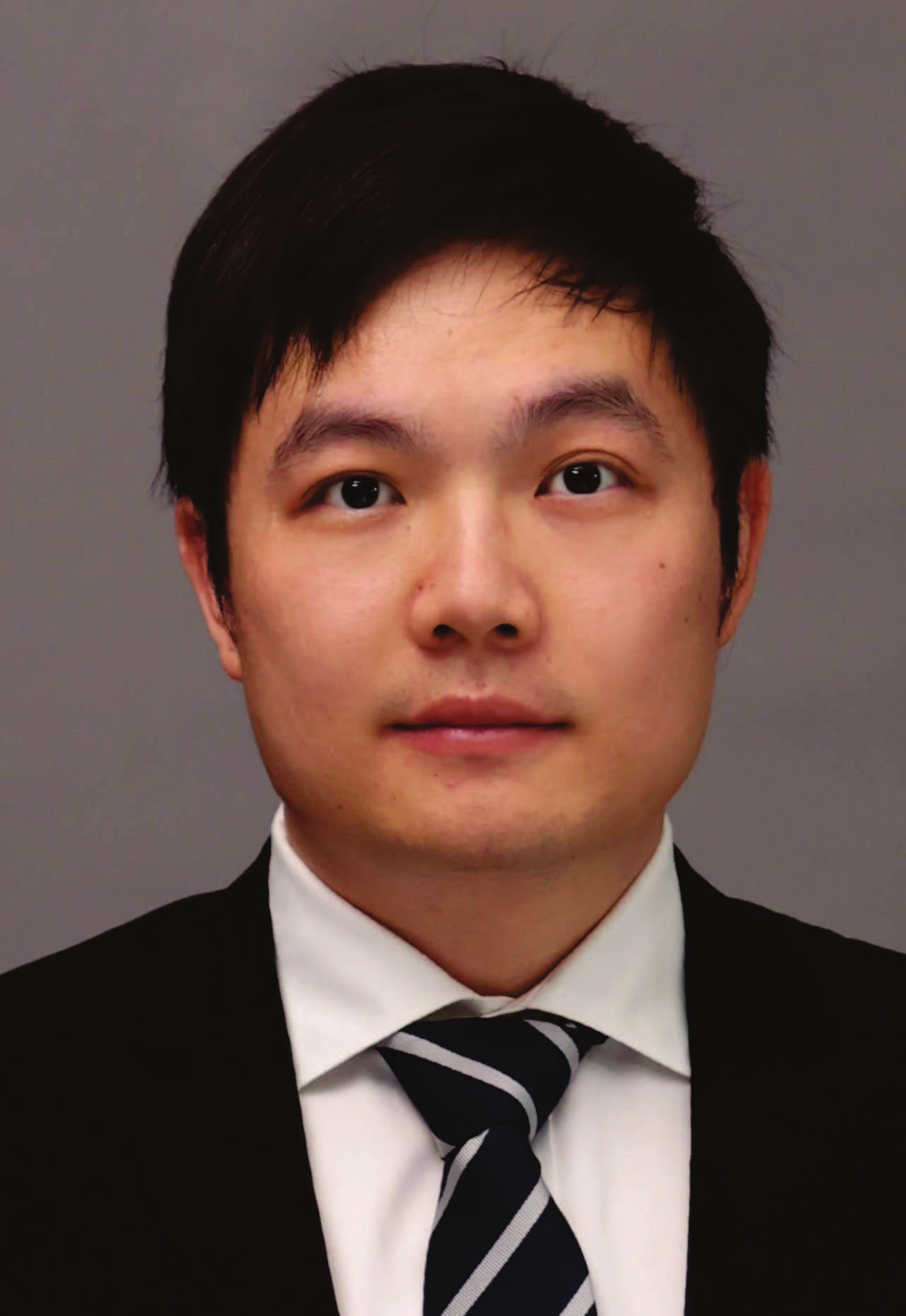}}]{Bo Qian}
(Member, IEEE) received the B.S. and M.S. degrees from the College of Mathematics at Sichuan University, Chengdu, China, in 2015 and 2018, respectively, and the Ph.D. degree in Information and Communication Engineering at Nanjing University, Nanjing, China, in 2022. From 2022 to 2024, he was a Postdoctoral Fellow with the Peng Cheng Laboratory, Shenzhen, China. From 2024 to 2026, he worked as a Researcher and Assistant Professor (Special Appointment) at the Information Systems Architecture Science Research Division, National Institute of Informatics, Tokyo, Japan. Currently, he is an Assistant Professor (Special Appointment) with the Graduate School of Information Science and Technology, The University of Tokyo, Japan. His research interests include cellular RAN, SAGIN, IoV, and Industrial Internet. He has published over 50 papers in leading journals and conferences, including IEEE JSAC, IEEE TMC, and IEEE INFOCOM, etc. He has won multiple best paper awards, including IEEE GLOBECOM, IEEE VTC-Fall, and WCSP, etc. He has served as the Guest Editor for some SCI Journals, Lead TPC Chair for IEEE INFOCOM 2026 DOICT-IndSoc Workshop, Co-Chair for IEEE/CIC ICCC 2025 AI-Native RAN Workshop, and many TPC members.
\end{IEEEbiography}

\begin{IEEEbiography}[{\includegraphics[width=1.25in,height=1.45in,clip,keepaspectratio]{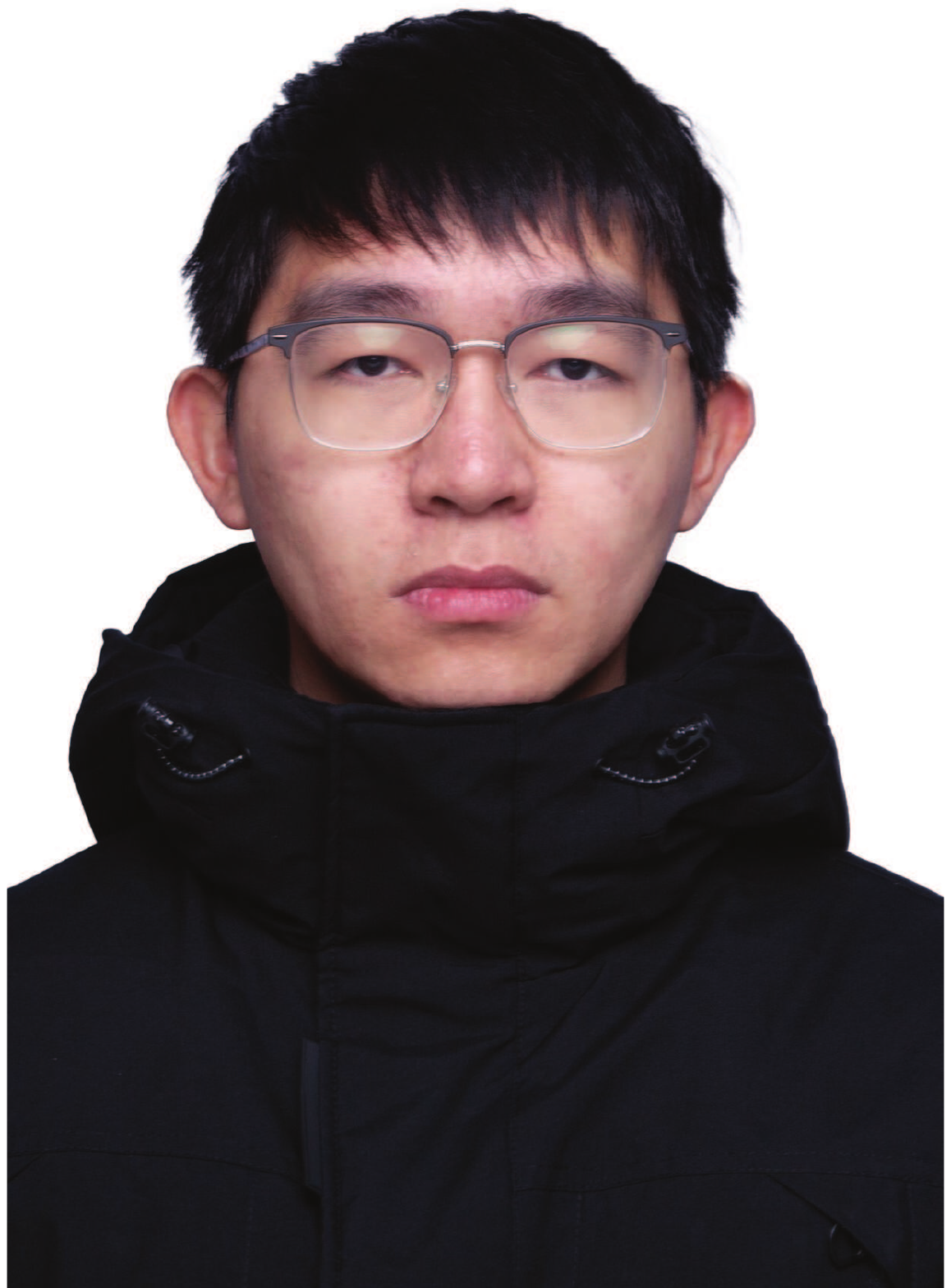}}]{Hanlin Wu}
(Student Member, IEEE) received the M.E. degree in electronic information science and technology in 2022 from NanJing University, Nanjing, China. He is currently pursuing the Ph.D degree with the school of information science and technology, The University of Tokyo, Tokyo, Japan. His research interests include autonomous driving, computer vision, intelligent transportation systems and vehicular networks.
\end{IEEEbiography}

\begin{IEEEbiography}[{\includegraphics[width=1.25in,height=1.35in,clip,keepaspectratio]{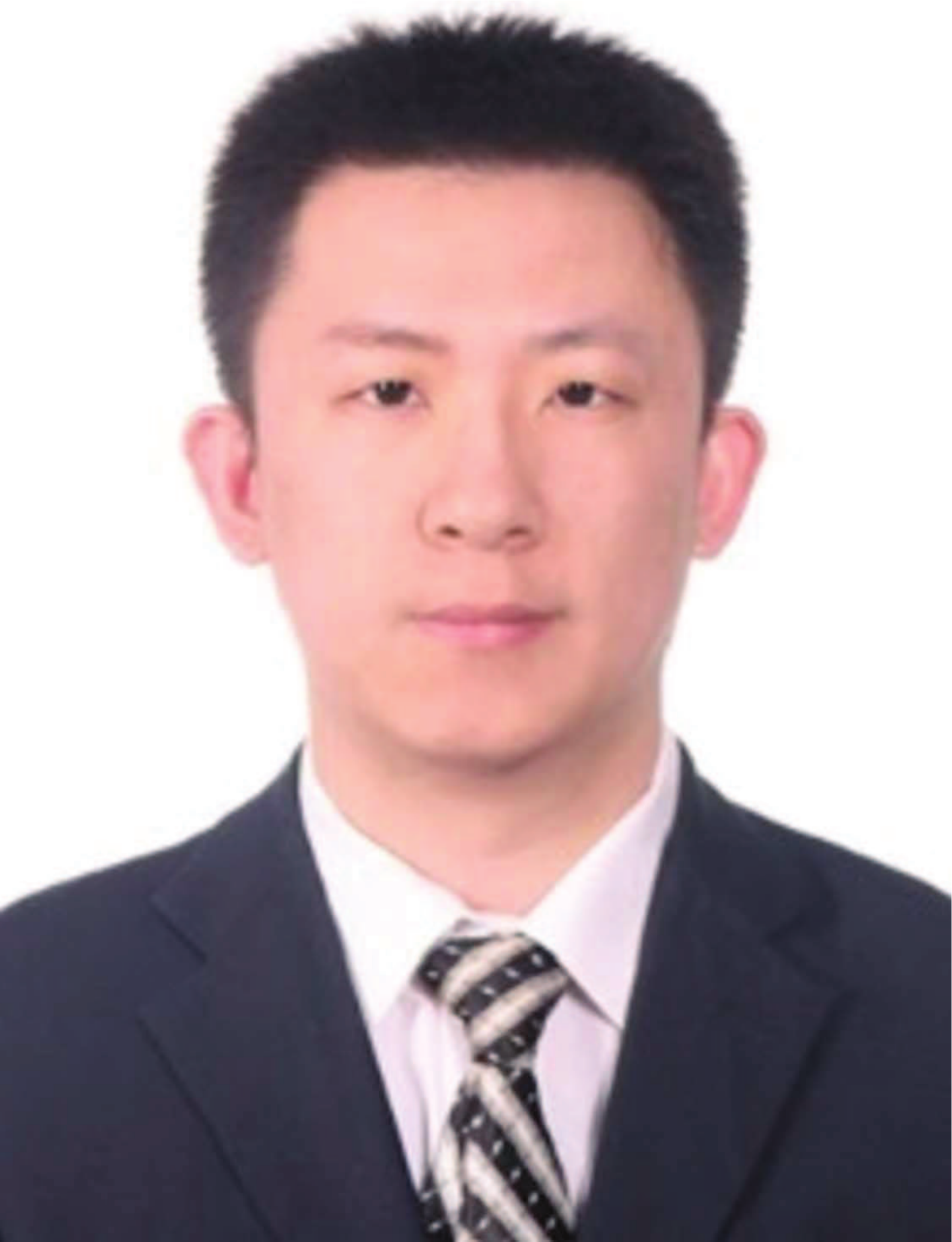}}]{Jiacheng Chen}
(Member, IEEE) received his Ph.D. degree in information and communications engineering from Shanghai Jiao Tong University, Shanghai, China, in 2018. From 2015 to 2016, he was a visiting scholar at BBCR group, University of Waterloo, Canada. Currently, he is an Associate Researcher in Peng Cheng Laboratory, Shenzhen, China. His research interests include 6G RAN and AI RAN. He has served as the guest editors for IEEE Internet of Things Journal (IoTJ) and Journal of Communications and Information Networks (JCIN), and the TPC Co-chair for IEEE INFOCOM 2026 DOICT-IndSoc Workshop. He has won the JCIN Best Paper Award, and the Chinese Institute of Electronics Outstanding Scientific Paper in the Field of Electronic Information.
\end{IEEEbiography}

\begin{IEEEbiography}[{\includegraphics[width=1.25in,height=1.35in,clip,keepaspectratio]{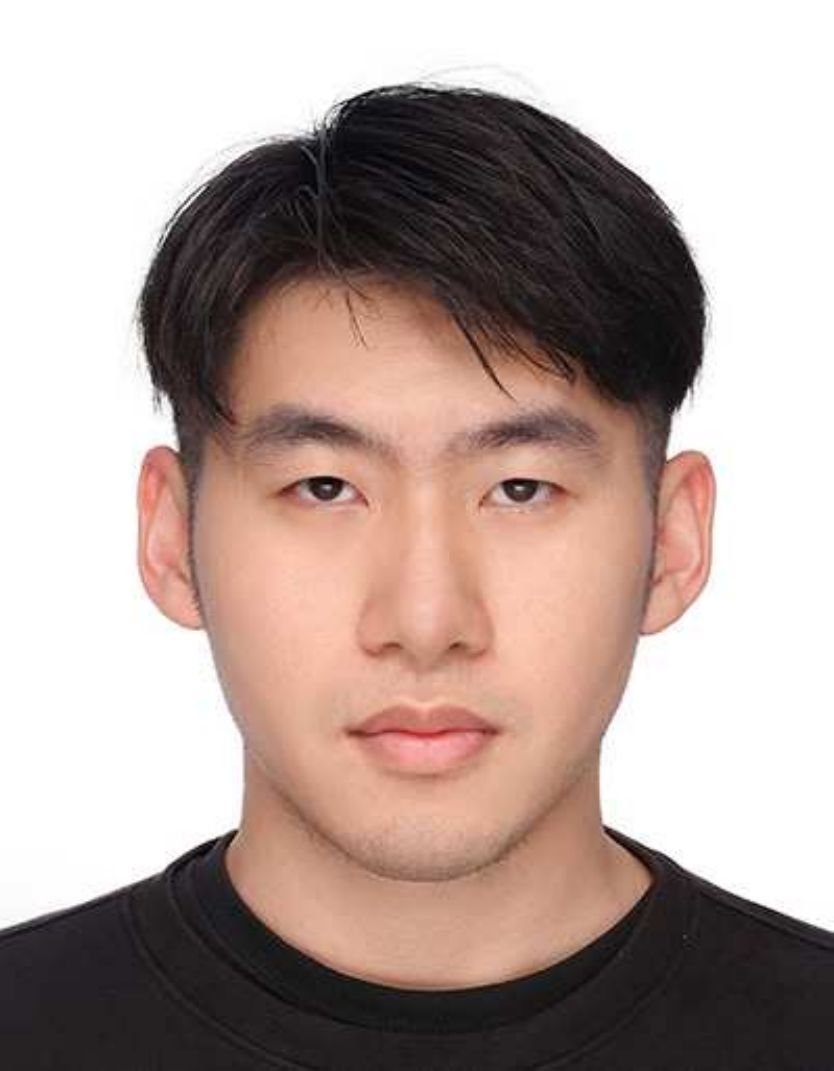}}]{Yunting Xu}
(Member, IEEE) received the PhD degree from Nanjing University in 2024. He is currently a Research Fellow with the College of Computing and Data Science (CCDS), Nanyang Technological University, Singapore. He mainly focuses on the artificial intelligence and networking optimization in the field of emerging wireless networks.
\end{IEEEbiography}

\begin{IEEEbiography}[{\includegraphics[width=1in,height=1.35in,clip,keepaspectratio]{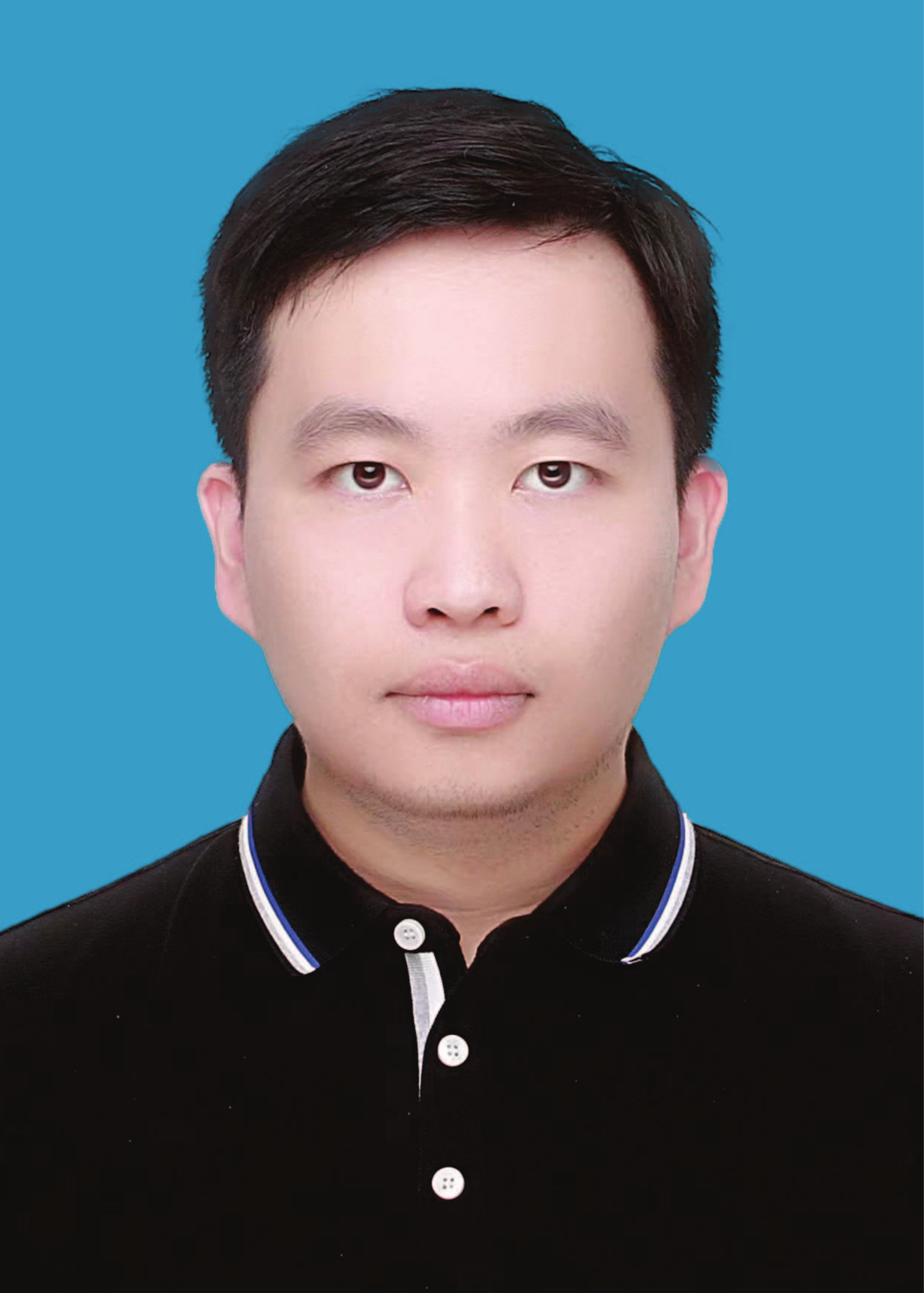}}]{Xiaoyu Wang}
received the M.S. degree in computer science from the University of Science and Technology of China, Hefei, China, in 2022. He is currently working toward the Ph.D. degree in informatics with the Graduate University for Advanced Studies, SOKENDAI, and the National Institute of Informatics, Tokyo, Japan. His research interests include recommender systems, large language models, and AI for networking.
\end{IEEEbiography}

\begin{IEEEbiography}[{\includegraphics[width=1.25in,height=1.35in,clip,keepaspectratio]{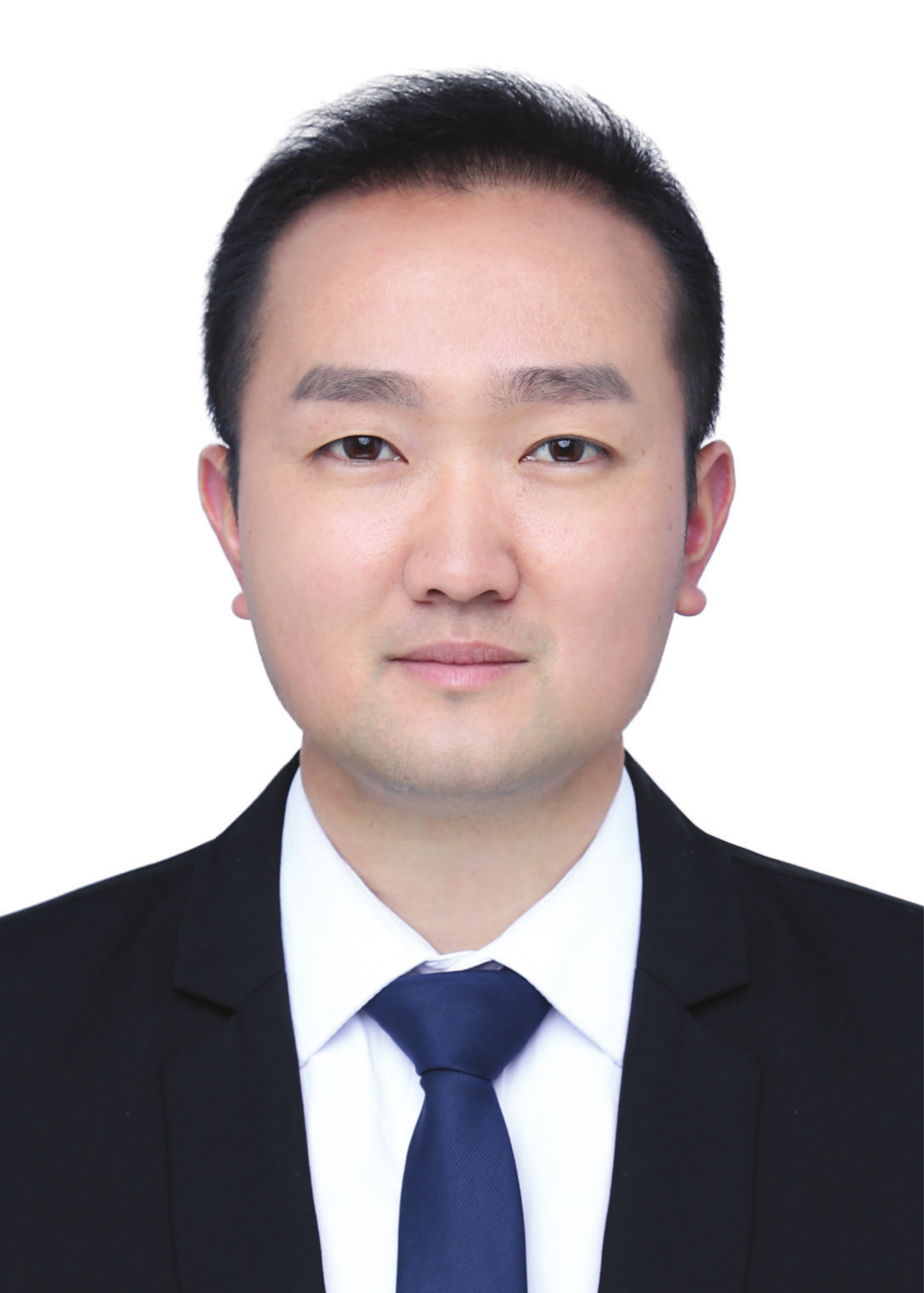}}]{Haibo Zhou}
(Fellow, IEEE) received the Ph.D. degree in information and communication engineering from Shanghai Jiao Tong University, Shanghai, China, in 2014. From 2014 to 2017, he was a Postdoctoral Fellow with the Broadband Communications Research Group, Department of Electrical and Computer Engineering, University of Waterloo. He is currently a Full Professor and Associate Dean with the School of Electronic Science and Engineering, Nanjing University, Nanjing, China. His research interests include resource management and protocol design in B5G/6G networks, vehicular ad hoc networks, and space-air-ground integrated networks. He has published over 150 scientific articles in leading journals and international conferences, including PIEEE, IEEE JSAC, IEEE TMC, and IEEE TWC etc. His research results have been highly visible, with 12 publications being ranked as ESI Highly Cited Papers. He has received multiple best paper awards, including IEEE GLOBECOM, IEEE VTC-Fall, IEEE PIMRC, and WCSP, etc. He was a recipient of the 2019 IEEE ComSoc Asia-Pacific Outstanding Young Researcher Award, 2023 IEEE ComSoc WTC Young Researcher Award, 2023-2024 IEEE ComSoc Distinguished Lecturer, and 2023-2025 IEEE VTS Distinguished Lecturer. He served as Track/Symposium Co-Chair for IEEE/CIC ICCC, IEEE VTC-Fall, WCSP, IEEE GLOBECOM, and IEEE ICC for many times, etc. He is currently an Associate Editor of the IEEE TWC, IEEE IoTJ, and IEEE TNSE, and IET Fellow and IEEE Fellow.
\end{IEEEbiography}

\begin{IEEEbiography}[{\includegraphics[width=1.25in,height=1.35in,clip,keepaspectratio]{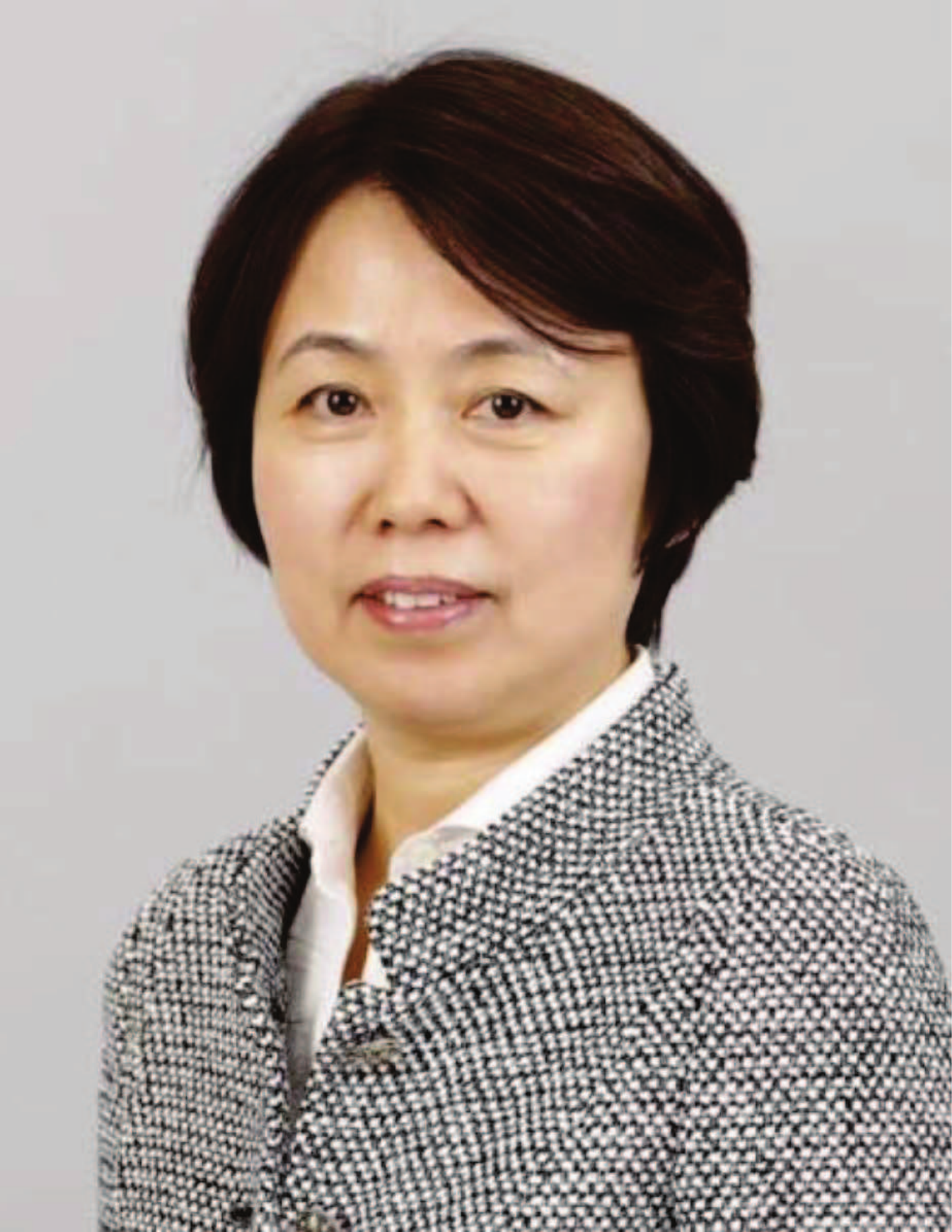}}]{Yusheng Ji}
(Fellow, IEEE) received the B.E., M.E., and Ph.D. degrees in electrical engineering from The University of Tokyo. She joined the National Center for Science Information Systems (NACSIS), Tokyo, Japan, in 1990, as an Assistant Professor and then became an Associate Professor. In 2000, she was an Associate Professor and then a Professor with the National Institute of Informatics (NII), Tokyo, and has been with the Graduate University for Advanced Studies, SOKENDAI, Japan, since 2002. She is a Professor Emeritus with NII and SOKENDAI since April 2026. She has published more than 600 refereed papers in the areas of network resource management, traffic control and performance evaluation, and mobile computing. She is a Distinguished Speaker of the IEEE Vehicular Technology Society. She was the recipient of numerous paper awards, including the IEEE Andrew P. Sage Best Transactions Paper Award and the IEEE Communications Society Outstanding Paper Award. She served as General Chair of IEEE INFOCOM 2026, TPC Chair of IEEE INFOCOM 2023, TPC Co-Chair, Symposium Co-Chair, and Track Co-Chair for IEEE ICC, GLOBECOM, and VTC, etc. She is/was an editor for transactions and magazine of IEEE, as well as IEICE and IPSJ.
\end{IEEEbiography}

\end{document}